\documentclass[12pt,draftclsnofoot,letterpaper,oneside,onecolumn]{IEEEtran}
\def\BibTeX{{\rm B\kern-.05em{\sc i\kern-.025em b}\kern-.08em
    T\kern-.1667em\lower.7ex\hbox{E}\kern-.125emX}}

\usepackage{amsmath}
\usepackage{amsthm}
\usepackage{amsfonts}
\usepackage{amssymb}
\usepackage{graphicx}
\usepackage{cite}
\usepackage{balance}
\usepackage{algorithm}
\usepackage{algorithmic}
\usepackage{enumerate}
\usepackage{color}
\usepackage{array}
\usepackage{indentfirst}
\usepackage{multirow}
\usepackage{booktabs}
\usepackage{color}
\usepackage{slashbox}
\usepackage{diagbox}
\usepackage{threeparttable}
\usepackage[svgnames,table]{xcolor}
\usepackage{bm}
\usepackage{bbm}
\usepackage{epstopdf}
\usepackage{subfig}

\makeatletter

\newcommand{\Rmnum}[1]{\expandafter\@slowromancap\romannumeral #1@}
\makeatother

\begin{document}

\title{ Time-Frequency Mask Aware Bi-directional LSTM: A Deep Learning Approach for Underwater Acoustic Signal Separation }

\author{\IEEEauthorblockN{Jie Chen, Chang Liu, Jiawu Xie, Jie An, and Nan Huang } 

\thanks{The authors are with the National Key Laboratory of Science and Technology on
Communications, University of Electronic Science and Technology of China, Chengdu 611731, China}

}

\maketitle

\begin{abstract}
The underwater acoustic signals separation is a key technique for the underwater communications. The existing methods are mostly model-based, and could not accurately characterise the practical underwater acoustic communication environment. They are only suitable for binary signal separation, but cannot handle multivariate signal separation. On the other hand, the recurrent neural network (RNN) shows powerful capability in extracting the features of the temporal sequences. Inspired by this, in this paper, we present a data-driven approach for underwater acoustic signals separation using deep learning technology. We use the Bi-directional Long Short-Term Memory (Bi-LSTM) to explore the features of Time-Frequency (T-F) mask, and propose a T-F mask aware Bi-LSTM for signal separation. Taking advantage of the sparseness of the T-F image, the designed Bi-LSTM network is able to extract the discriminative features for separation, which further improves the separation performance. In particular, this method breaks through the limitations of the existing methods, not only achieves good results in multivariate separation, but also effectively separates signals when mixed with 40dB Gaussian noise signals. The experimental results show that this method can achieve a $97\%$ guarantee ratio (PSR), and the average similarity coefficient of the multivariate signal separation is stable above 0.8 under high noise conditions.
\end{abstract}


\clearpage

\section{Introduction}
\label{sec:introduction}
At present, underwater acoustic communication \cite{chen2021underwater} mainly uses sonar technology to detect, locate and identify the underwater targets. However, the sonar technology needs to overcome the noises like ship noise and ocean noise \cite{kim2016underwater, liu2017biologically, bereketli2012remotely}. Source separation technology is a good way to reduce the noises \cite{rahmati2017unisec, cardoso1996independent, wang2009research,cardoso1998blind}, which attracts tremendous research from both academia and industry.
Among these source separation methods, the blind source separation (BSS) is a classical method\cite{comon2010handbook, he2009application, kirsteins2003blind}, which consists of mathematical model, objective function, separation algorithm and evaluation criteria \cite{heli2011localization, ozerov2009multichannel}.
During the research of the BSS algorithm, two approaches are always studied and employed. One is based on independent component analysis (ICA) \cite{comon1994independent} which works well when the number of sources $N$ is less than or equal to the number of sensors $M$. The use of ICA is not limited to linear instantaneous mixing, it is also used to solve the separation problem of convolutional mixing and even nonlinear mixing. Another relies on the sparseness of source signals which works well when $N$ is more than $M$, like binary T-F mask approach \cite{jourjine2000blind}. The binary T-F mask approach extracts a signal by calculating a binary masking matrix of the signal. It has the advantage of real-time, and in recent years it has also been applied to underwater acoustic separation in combination with underwater sound characteristics.

In view of underdetermination in underwater acoustic communication, this paper studies the method of binary time-frequency mask based on sparsity. The traditional binary T-F mask method chooses features which are performed manually, by using the observation signals. Due to the outliers and distribution of anisotropic variance, the traditional feature extraction method has certain limitations: it can only be used in binary signal separation, but the effect is poor in multiple signal separation, which cannot meet the requirements of separation accuracy. At present, the improvement of binary T-F masking method still stays in feature design \cite{yilmaz2004blind, araki2007underdetermined, Araki2004Underdetermined}. However, it is not easy for human experts to design good features. These artificial features are easily affected by outlier problems and have strict requirements on the selection of source location. As an alternative, on top of the traditional binary T-F masking, the method of extracting the original features of the underwater acoustic source using the deep neural network has shown good performance. At present, this method has been used to solve image recognition, natural language processing (NLP) and even communication problems \cite{tian2014learning}. Deep learning approach \cite{liu2021learning, liu2021edge, lxm2020deepresidual, liu2020deepresidual} also makes a breakthrough in the separation of signals. Therefore, we extract the features of the underwater acoustic signals by means of deep learning approach. The main contribution of this work are as follows:

 \noindent
 \hangafter=1
 \setlength{\hangindent}{2em}
 (1) We propose a deep learning method based on Bi-LSTM. This method uses the powerful feature extraction capability of RNN, not only improves the performance of separating binary signals, but also achieves good results in ternary or multivariate signal separation experiments. This overcomes the limitations of the previous separation of single targets from deep learning sources.

 \noindent
 \hangafter=1
 \setlength{\hangindent}{2em}
 (2) We improved the training sample with the idea of embedding: embedding each T-F point into a high-dimensional space so that each T-F point can be represented as a vector, and then adding energy-based reference labels to the training sample. This makes the T-F points of different sources more distinct and clustering easier in the process of neural network learning.

 \noindent
 \hangafter=1
 \setlength{\hangindent}{2em}
 (3) We have carried out a lot of experiments on the separation performance of this method by using the unknown noise generated randomly and the marine noise actually collected. The experimental results show that this method can effectively separate the noise as long as the number of clustering $K$ is increased. It is proved that this method still has good robustness and scalability in the actual Marine environment with sufficient complexity.

The rest of the paper is organized as follows. In Section II, we introduce traditional system model of the underwater acoustic source separation. Then, in Section III, we present the proposed approach description, including offline training and online test. Section IV presents the experiments. Finally, conclusion is drawn in Section V.

\section{Mainstream Method: Binary Time - Frequency Masking Method}
The binary T-F mask approach separates the underwater acoustic signals according to the auditory masking, using the underwater acoustic source which dominates the energy in a certain T-F domain. Although the target signals received by the system have the varying degrees of frequency band overlap, the main energy of different target signals is usually hidden in different frequency bands. So, the binary mask approach can use this property to realize underwater acoustic signal separation by clustering the T-F bins. To cluster such T-F bins, the traditional method uses the observation signals and calculates manually to gets features.

\subsection{Restrictions On Using Existing Methods}
The use of binary T-F masking techniques must satisfy the sparsity condition. Since the sound signal is generally not sparse in the time domain, it needs to be transformed into the T-F domain by some transformation. However, in the actual separation process of underwater acoustic signals, it is found that the energy of different underwater acoustic radiation signals is usually concentrated in different frequency bands, and the target radiation signals received by the system will have different frequency band aliasing phenomena. The study found that as long as the underwater acoustic signal can satisfy the absolute dominant condition of energy, the binary T-F masking algorithm can be used to achieve separation. This condition is written as:
\begin{equation}
|X_i(t,f)|\gg|X_j(t,f)|, i\neq j,\forall t,f ,
\end{equation}where $X_i(t,f)$ is the Short Time Fourier Transform (STFT) of signal $x_i(t)$. By using STFT, the signals in the time domain can be transformed to the T-F domain which can satisfy the property of sparsity. Geometric features for clustering are calculated based on this constraint.

This condition can also be understood as the fact that the overlap of the T-F domain is a relatively small portion of one of the underwater acoustic signals, so that ignoring the information of this part does not affect the recovery of the entire signal.

\begin{figure}[!t]
	\centering
	\includegraphics[width=3.5 in]{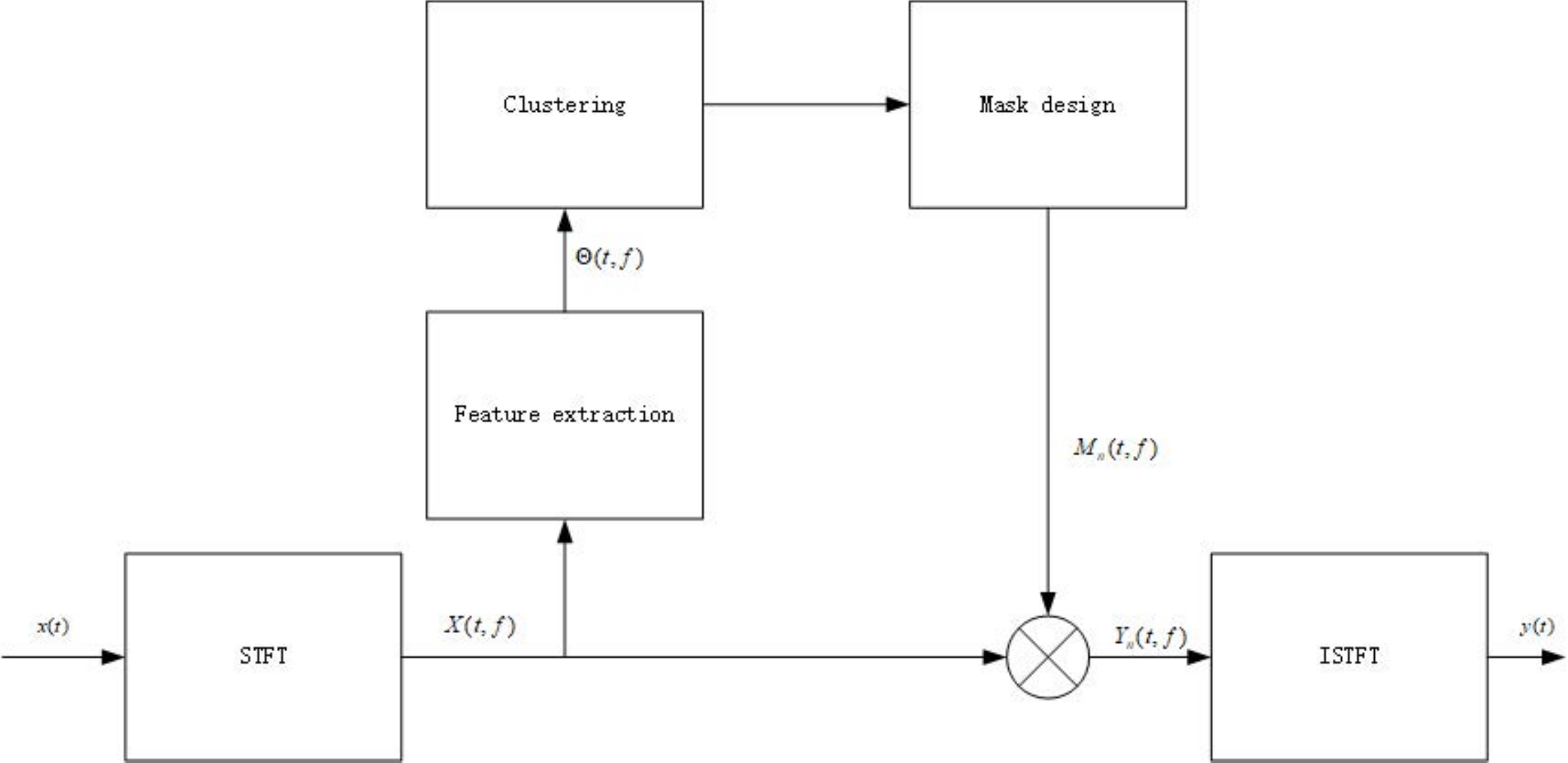}
	\caption{Example of T-F mask approach}
	\label{TF}
\end{figure}

\subsection{Signal Separation Steps In Underdetermined Case}
This approach can be summarized in Fig.\ref{TF} .Based on the sparsity condition of absolute dominance of energy, in the case of underdetermined, the idea of using the binary T-F masking method for water acoustic blind separation is as follows:

(1) STFT. Let the sampling frequency of the observation signal be $ f_s $, and convert the time domain signal $x(t)$ into the T-F domain representation by using the T-point STFT transform:
\begin{equation}\label{2-35}
X(t,f)=\sum_{r=-T/2}^{T/2-1} x(r+tL)win(r)e^{-j2\pi fr}
\end{equation}where $t$ is the time frame index, $f$ is the frequency point, $T$ is the length of the window, and $L$ is the moving length of the window. $win(r)$ represents the window function. Commonly used are rectangular window, Hanning window and Hamming window. In the subsequent Inverse Short Time Fourier Transform (ISTFT), we used the Hanning window to transform to ensure consistent parameters.

(2) Feature extraction. The source signal $X(t,f)$ satisfying the sparse condition is obtained by STFT transform, and the feature vector $\Theta (t,f)$ is calculated therefrom. On this eigenvector, there are differences between different sources, which can be measured by distance. The $\Theta (t,f)$ is generally composed of the geometric characteristic magnitude $\alpha (t,f)$ and the phase difference $\phi (t,f)$ between the observed signals.

Taking two observation signals $X_1(t,f)$ $X_2(t,f)$ as an example, the eigenvector $\Theta (t,f)$, the order of magnitude $\alpha (t,f)$, and the phase difference $\phi (t,f)$ can be calculated by the following equations:


\begin{equation}
\Theta(t, f)=\left[\alpha (t,f), \phi (t,f)\right]
\end{equation}
\begin{equation}
\alpha(t, f)=\frac{|X_2(t,f)|}{|X_1(t,f)|}
\end{equation}
\begin{equation}
\phi(t,f)=\arg \left(\frac{X_2(t,f)}{X_1(t,f)}\right)
\end{equation}

The phase difference is usually normalized to avoid frequency sequencing problems, and the above equation can be written as.
\begin{equation}
\phi(t,f)=\frac{1}{2\pi f}\arg \left(\frac{X_2(t,f)}{X_1(t,f)}\right)
\end{equation}

Expanded to the case where there are multiple observation signals, the order of magnitude $\alpha(t,f)$ and the phase difference $\phi(t,f)$ are expressed as:
\begin{equation}
\label{2-43}
\alpha(t, f)=\left[\frac{\left|X_{1}(t, f)\right|}{A(t, f)}, \ldots \frac{\left|X_{n}(t, f)\right|}{A(t, f)}\right]
\end{equation}
\begin{equation}
A(t, f)=\sqrt{\sum_{j=1}^{n}\left|X_{j}(t, f)\right|^{2}}
\end{equation}
\begin{equation}
\label{2-45}
\phi(t, f)=\left[\frac{1}{\beta_{1} f} \arg \left(\frac{X_{1}(t, f)}{X_{B}(t, f)}\right), \ldots, \frac{1}{\beta_{n} f} \arg \left(\frac{X_{n}(t, f)}{X_{B}(t, f)}\right)\right]
\end{equation}where, $A(t,f)$ is the normalization coefficient of order of magnitude; $\beta_j=\beta=4\pi d_{max}/c, j=1,...,n$ is the weight coefficient of phase difference, subscript $B$ represents the label of the reference observation signal, $c$ represents the sound propagation speed, and $d_{max}$ represents the maximum distance between the reference observation signal and other observation signals.

Express $\Theta (t,f)$ as a plural form with the following equation:
\begin{equation}
\tilde{\Theta}_{i}(t, f)=\left|X_{i}(t, f)\right| \exp \left[j \frac{\arg \left(X_{i}(t, f) / X_{B}(t, f)\right)}{\beta_{i} f}\right]
\end{equation}

Normalization of the above equation yields a eigenvector representation of the multi-observed signal:
\begin{equation}
\Theta_{i}(t, f)=\tilde{\Theta}_{i}(t, f) /\left\|\tilde{\Theta}_{i}(t, f)\right\|
\end{equation}
\begin{equation}
\mathbf{\Theta}(t, f)=\left[\Theta_{1}(t, f), \ldots \Theta_{n}(t, f)\right]^{\mathrm{T}}
\end{equation}\textbf{}

From the equations, we can know that $\mathbf{\Theta}(t, f)$ which is the features extracted by us are influenced by all kinds of aspects.

(3) Cluster analysis. Clustering the feature vector $\mathbf{\Theta}(t, f)$ can obtain m clusters $C_1,...,C_m$ corresponding to m source signals. Past clustering methods have manual clustering \cite{Jourjine2002Blind}, kernel density estimation \cite{Roman2003Speech}, or maximum likelihood(ML) based gradient search method \cite{rickard2001real}. Because K-means clustering has the characteristics of simple, convenient and fast convergence, it has become the most commonly used method for cluster analysis. K-means can minimize the sum $\Upsilon$ of the Euclidean Distances (ED) of each source signal and the corresponding cluster center $c_k$, and automatically divide the samples into m clusters. The equation is expressed as:
\begin{equation}\label{sum11}
\Upsilon=\sum_{k=1}^{m}{\Upsilon_k}
\end{equation}
\begin{equation}\label{sum12}
\Upsilon_k = \sum_{\mathbf{\Theta}(t,f)\in C_k}\|\mathbf{\Theta}(t,f)-c_k\|^2
\end{equation}

First, m cluster centers $c_1,c_2,...,c_m$ are randomly initialized, each feature vector is assigned by iterative equation \eqref{2-51}. Then the feature vector $\Theta(t, f)$ closest to the mean vector $c_k$ is found and assigned as a cluster:
\begin{equation}\label{2-51}
C_k=\{\Theta (t,f)|k=\underset{k}{argmin}\left \| \Theta(t,f)-c_k \right \|^2\}
\end{equation}

Calculate the mean of all feature vectors belonging to $c_k$ and correct the cluster center:
\begin{equation}
c_k\leftarrow E[\mathbf{\Theta}(t,f)]_{\mathbf{\Theta}\in C_k}
\end{equation}

Substituting the updated mean vector into the equations \eqref{sum11} and \eqref{sum12} calculates the objective function $\Upsilon$. If $\Upsilon$ converges, then the set $C_k, k =1,2,...,m$ corresponding to each source is obtained after the iteration ends.

(4) Binary T-F masking. Using the results obtained by clustering, a binary T-F masking matrix is constructed. The binary T-F masking matrix is a matrix consisting of 0 and 1 values and whose size is consistent with the T-F matrix. This is similar to the binary test in spectrum sensing \cite{liu2019maximum, xie2020deep, liu2014blind, liu2016blind}. The matrix sets the mask value to 1 or 0 according to whether each T-F point belongs to the target signal, indicating whether the T-F point information belongs to the source signal.
$$
M_k(t,f)=
\begin{cases}
1,  & \mathbf{\Theta (t,f)}\in C_k \\
0,  & \text{others}
\end{cases}
$$

Substituting the following equation gives the spectrum of the estimated signal:
\begin{equation}\label{yk(t,f)}
Y_k(t,f)=M_k (t,f)X(t,f)
\end{equation}

(5) Inverse Short-Time Fourier Transform (ISTFT). After obtaining the T-F domain estimation, the final step needs to complete the recovery of the time domain signal $y_k(t)$ using ISTFT and overlap retention method \cite{raki2005reducing}:
\begin{equation}\label{yk(t)}
y_k(t)=\frac{1}{A}\sum_{l=0}^{L-1 }y_{k}^{d+l}(t)
\end{equation}

When using ISTFT, the parameters need to be the same as those of STFT using equation \eqref{2-35}. Where A is a constant, related to the window function, $A=0.5T/L$ when using Hanning window, and $y_{k}^{d}(t)$ is expressed as follows:
$$
{y_{k}^{m}(t)=
\begin{cases}
\sum_{f\in {0,\frac{1}{T}f_s,...,\frac{T-1}{T}f_s}}&Y_k(m,f)e^{j2\pi fr}, \\&mL\leq t\leq mL+T-1 \\
0,  &\text{others}
\end{cases}
}
$$where, $r=t-mL$.

\subsection{Evaluation Of Separation Performance}
In order to verify the separation performance of the algorithm after adding noise, we simulated the binary time-frequency masking method. The T-F masking method requires the signal to meet the conditions of WDO or energy dominance. Therefore, the LFM signal is selected for simulation to facilitate the aliasing operation of the signal at time and frequency. The detailed experimental process is described in Section IV. The experimental results show that when there is no noise, each signal can be well recovered, and the method can correctly divide the T-F region of each signal. Once noise is added, performance deteriorates. The estimated masking matrix not only loses some information of the signal itself, but also receives the T-F information of other signals.

\section{Proposed Method}
\begin{figure}[!t]
	\centering
	\includegraphics[width=3.5 in]{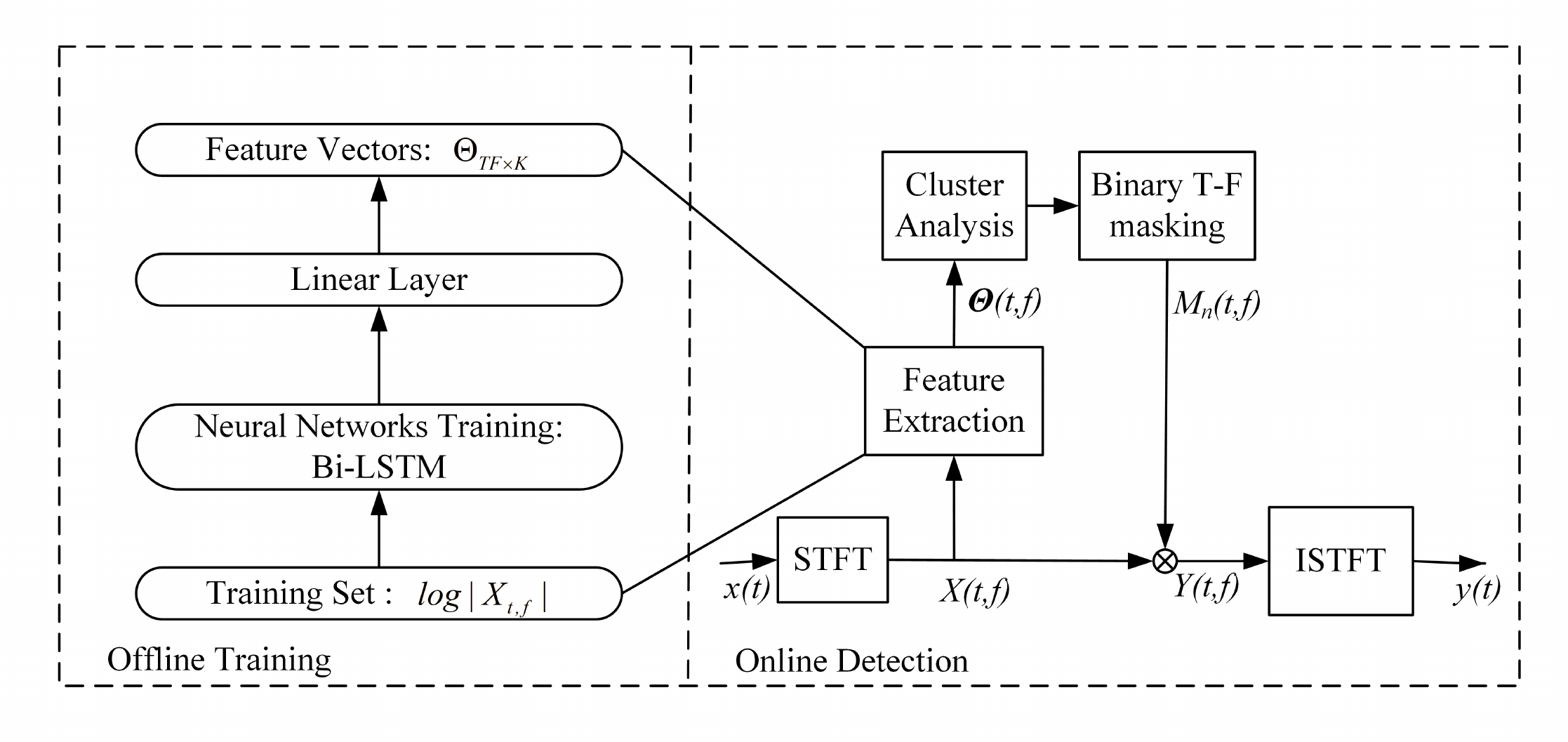}	
	\caption{Framework of proposed approach}
	\label{archi}
\end{figure}
In recent years, deep learning has been successfully applied in speech separation\cite{inproceedings, Huang2014Deep, Huang2015Joint}, and these previous attempts have generally assumed that the numbers and types of sources are fixed. However, in the case of underwater acoustic signal separation, we have to consider two problems: 1) the model can be used to separate arbitrary kinds of underwater acoustic sources, i.e. generalization problem; 2) the model can be used to separate arbitrary numbers of underwater acoustic sources, i.e. scalability problem. Unlike previous attempts and in this article, we use deep learning methods to learn a mapping for the input that is amenable to clustering, and it is helpful to overcome the above two shortcomings. The architecture of the proposed method is illustrated in Fig.\ref{archi}.

\begin{figure*}[ht]
	\centering
	\includegraphics[width=\linewidth]{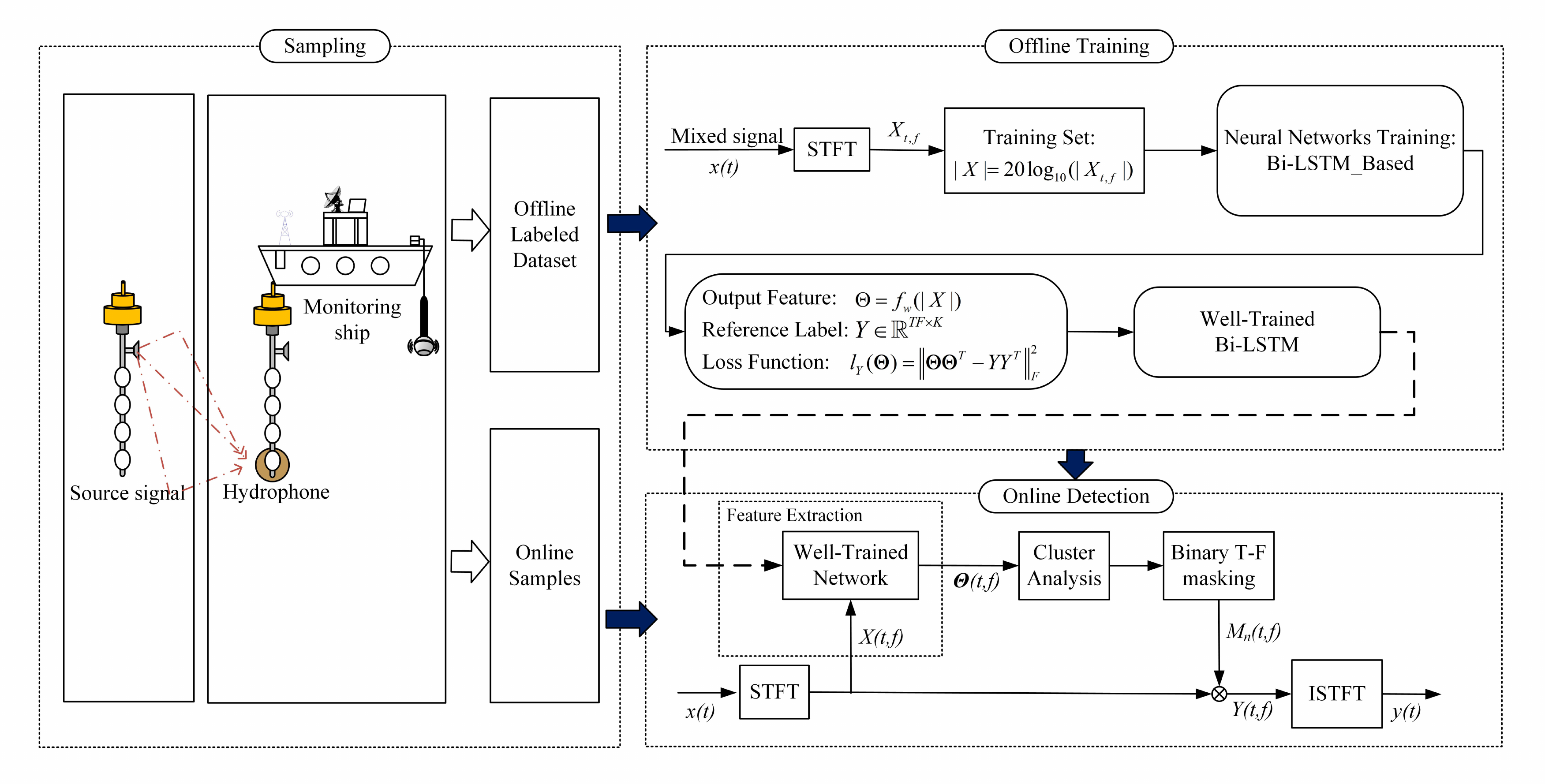}
	\caption{Flow chart of proposed method for underwater acoustic signal separation}
	\label{flow}
\end{figure*}

Based on the traditional binary T-F masking method, this scheme uses the deep neural network to extract features from the original underwater acoustic data instead of artificial feature extraction. The program is divided into two stages: offline training and online testing. (1) Offline training phase. The data is obtained from the measured underwater sound database and processed to obtain training samples. The T-F map of the underwater acoustic signal is obtained mainly through STFT, and then the neural network is trained. In order to make the network learn from the original underwater acoustic characteristics to cluster-oriented features, this paper sets the appropriate objective function to make the characteristics of the network output easier to cluster; (2) online testing phase. The artificial feature extraction method in the traditional binary T-F masking method is replaced by the network with the previous stage learning, and the different mixed water acoustic signals are used to test whether the separation performance of the scheme meets the requirements. The specific method flow chart is shown as Fig.\ref{flow}.

\subsection{Feature Extraction Based on Deep Neural Network}

In fact, in order to achieve good separation performance after clustering, it is required that the clustering features have good distinguishing characteristics. In recent years, many studies have used deep neural networks \cite{liu2020location} to obtain powerful characterizations for clustering \cite{tian2014learning, xie2020deep, xie2020unsupervised, huang2014deep1, mikolov2013distributed, song2013auto, alqahtani2018deep, liu2020deeptransfer, liu2019deep}, which have achieved good results in image recognition and NLP. They are characterized by embedding the original data features into the new feature space, making the transformed features more suitable for clustering.

In addition to the target underwater acoustic signal, there are also ship radiated noise and Marine environment noise in the sonar system. Due to varying degrees of decay in the ocean, the main energy of these noises is concentrated in different frequencies. The main sound source frequencies are shown in table \ref{sound_source}
\begin{table}[!t]
	\caption{Major Ocean Sound Source Frequency}
	\label{sound_source}
	\centering
	\begin{tabular}{|c|c|}
		\hline
		Sound source              & Frequency       \\ \hline
		Marine life               & 0.5 - 20 KHz    \\ \hline
		Radiated noise from ships & less than 1 kHz \\ \hline
		Surface ships             & 100 - 500 Hz    \\ \hline
		Submarines                & 100 - 500 Hz    \\ \hline
		Torpedoes                 & 500 - 1500 Hz   \\ \hline
	\end{tabular}
	
\end{table}


For a communication sonar, to receive transmitted signals from other sonar platforms, the receiving bandwidth of the receiver is about 100Hz to 3000Hz, and the receiver has prior knowledge of these detection signals \cite{tian2014learning}. According to the characteristics of underwater acoustic signals, if using neural network to "divide" different types of signals in the water audio frequency domain, and then using Fourier transform signal processing method to restore the signal, finally the target signal can be separated.

According to the embedding principle, the role of the deep neural network used in this section is to map the original features (immediate frequency features) of the measured data to the new feature space. Each T-F point is converted into a vector. Each vector has a different position in the new feature space, depending on the amount of energy at the T-F point. These vectors are then "divided" into a number of reasonable ranges based on the distance between the vectors. That is, the T-F vectors belonging to the same underwater sound source have similarities such that the distance is the smallest, and the T-F vectors belonging to different underwater sound sources have a large distance. Finally, they can be easily divided by a simple clustering algorithm.

Suppose a mixed water acoustic signal is transformed by STFT to obtain the original T-F characteristic $X_{t,f}\in R^{T*F}$, where $t$ is the number of time frames and $f$ is the frequency point. Taking the logarithmic amplitude spectrum $20log10(|X_{t,f}|)$ as the input of the network, for the convenience of description, the latter is uniformly recorded as $|X|$. $|X|$ can also be regarded as a sequence $[\chi_1,\chi _2,...,\chi _T]$ composed of spectral information $\chi _i\in R^F$ over a plurality of consecutive times. The deep neural network is parameterized by $\omega$, and the features generated based on the network are expressed as:
\begin{equation}
\Theta=f_{\omega}(|X|)
\end{equation}Among them, $\Theta=[\theta_1,\theta_2,...,\theta_{TF}]^T \in R^{TF*K}$ is the whole amplitude information $|X|$ of the underwater acoustic signal. The cluster-oriented $K$-dimensional embedding feature learned by neural network. During the training process, the network sequentially maps the spectrum information $\chi_i$ on each time step to a new feature space, and finally outputs it as an $F*K$-dimensional vector. This can be considered as encoding each T-F point in the original T-F feature $\chi_i$, and each T-F point after encoding is represented by a row vector $\theta_j$ of dimension $K$. Here $\theta_i$ is the unit vector, i.e. $|\theta_j|^2=1$.

The goal of training is to allow the line vector of the network output feature $\Theta$ to be divided into different water sources. That is, $\theta_j$ satisfies the vector distances belonging to the same water source, and the vectors belonging to different water sources are far away, so as to achieve the purpose of separating the underwater sound.

Assuming that there is a mixed underwater sound source in the water area, it is composed of C kinds of underwater sound sources:
\begin{equation}\label{3-2}
x(t)=\alpha_1s_1(t)+\alpha_2s_2(t)+...+\alpha_C s_C(t)
\end{equation}

Before sending mixed signals to network training, compare the energy of each source signal at each time and frequency point. First, set the reference label $Y\in R^{TF*C}$ to divide the time and frequency points, and compare the energy of these C kinds of underwater sound sources at various time and frequency points. The energy-dominated underwater sound source will mark the time and frequency points. For example, the energy of the no. c ($c\in {1,2,...,C}$) underwater sound, dominates at the no. n ($n\in {1,2,...,TF}$) time and frequency points, then $y_{n,c}=1$. Therefore, the loss function of the model can be set as:

\begin{align}
l_Y(\Theta)&=\|\Theta\Theta^T-YY^T\|^2_{F} \nonumber\\&=\sum_{i,j}(\langle \theta_i,\theta_j\rangle-\langle y_i,y_j\rangle)^2 \nonumber \\
&=\sum_{i,j:y_i=y_j}(|\theta_i-\theta_j|^2-1)+\sum_{i,j:y_i\neq y_j}\langle \theta_i, \theta_j \rangle^2
\end{align}

where $\|\bullet\|_F^2$is the squared Frobenius norm \cite{Hershey2016Deep}. In the process of minimizing the loss function, the two vectors divided in the same water source will be closer and closer, and the distance between the two vectors divided under different water sources will be farther and farther. At the same time, since $(YP)(YP)^T=YY^T$ exists for any permutation matrix P, the method can ensure that the label arrangement and the number of all training samples are independent.

\subsection{Offline Training: Test Network Design Based on RNN, LSTM and Bi-LSTM Respectively}
$\textbf{Input and reference label processing}$: First, randomly take [2 $C$] underwater acoustic audio files from the file library and mix them according to equation \eqref{3-2}. Each audio file needs to be averaged before entering the network training:
\begin{equation}
{s}'(t)=s(t)-E[s(t)]
\end{equation}
\begin{equation}
{s}''(t)=\frac{{s}'(t)}{max(|{s}'(t)|)}
\end{equation}The mixing coefficient $\alpha$ is randomly taken as an arbitrary number of the16 interval [3/4, 1].

According to equation \eqref{2-35}, the mixed signal is 32 ms window length, 8 ms time shift STFT, and the log amplitude spectrum X is taken. For a 16 s audio, it can be split into 500 samples of size 706. At the same time, take logarithmic amplitude spectrum of each source signal that makes up the mixed signal, and compare the magnitude of energy at each time and frequency point together to form the reference label Y with the same shape as X. To ensure local accuracy, each iteration consists of a sequence of time steps from multiple input samples of X and Y, and each sequence is 50\% overlapped to form a minimum batch-pair network for training.

In the offline training phase, in order to more clearly introduce the proposed Bi-LSTM structure of this paper and highlight its superiority with other neural networks, we tested it based on three structures: RNN, LSTM and Bi-LSTM. In addition, since LSTM is closely related to Bi-LSTM, the following section will first give a brief description of the adopted LSTM structure, followed by a detailed introduction of Bi-LSTM.

\textbf{Structure 1(LSTM-based):} RNN has long-term dependency problems. As the structural model of RNN gets deeper, RNN needs to repeatedly apply the same operations to each moment in the long-term sequence to generate a very deep computational graph. Coupled with model parameter sharing, RNN is prone to losing the ability to learn previous information, making optimization extremely difficult. Unlike RNN's regular loop body structure, LSTM uses neurons dedicated to memory storage. The neuron is a special network structure with three "gate" structures, which are called input gates, output gates, and forgetting gates. During training, the LSTM relies on these gated operations (reset and read and write operations) to selectively influence the state of each moment in the network.

After the investigation, we know that feature extraction can be performed using RNN. In particular, we use LSTM networks which is an improvement of RNN in this work\cite{Gers2002Learning}.
\begin{figure}[!t]
	\centering
	\includegraphics[width=3.5 in]{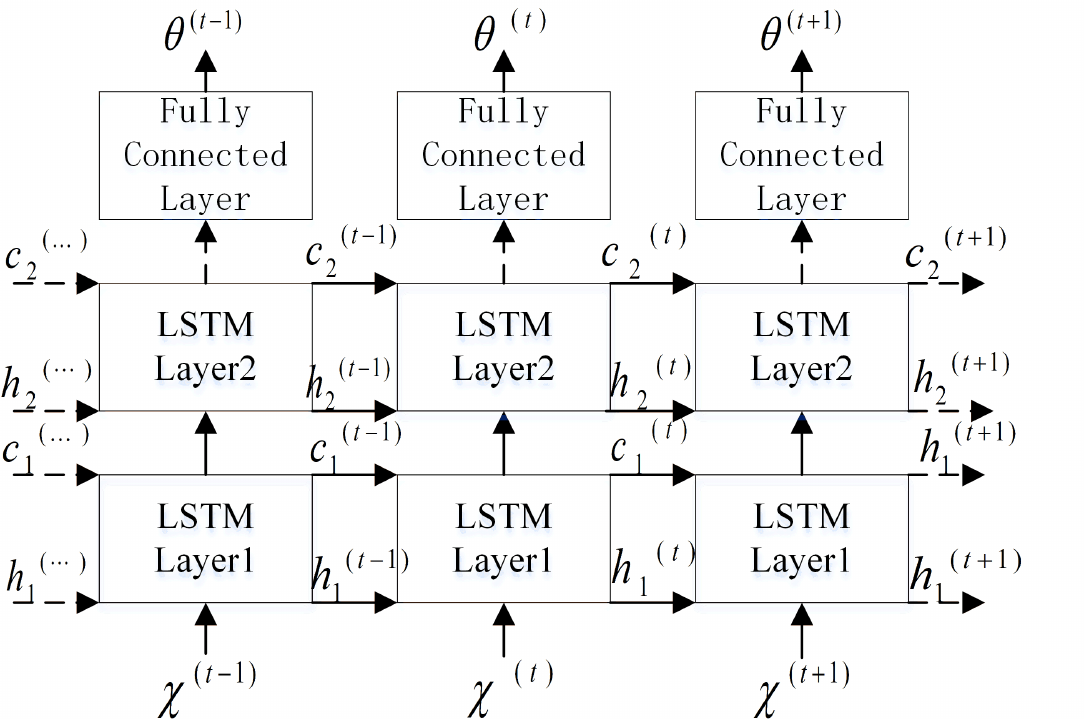}
	\caption{LSTM underwater acoustic separation network structure}
	\label{LSTM}
\end{figure}
LSTM can form a deep LSTM network by stacking, repeating the loop body at each moment to enhance the expressive ability of the model. The parameters of the loop body of each layer are the same, and the loop body parameters of different levels can be different. A schematic diagram of the network structure for water acoustic separation using multi-layer LSTM is shown in Fig.\ref{LSTM}. By stacking, the neural network can learn deeper expressions and finally embed them into the K-dimensional features.

\textbf{Structure 2(Bi-LSTM-based):} The transmission of the two network structures, RNN and LSTM, is one-way from front to back, that is, the state at time $t$ can only capture information from the past sequence $x^{1},...,x^{t-1}$ and the current input $x^{t}$. For some problems, however, the prediction of the output may depend on the entire sequence. For example, in speech recognition, some words currently have multiple interpretations, which need to be judged in combination with context and context. Therefore, the processing of the voice needs to refer to the pronunciation information of the past and the future in order to have a more accurate effect.

It is also possible to encounter the same problem in the field of underwater sound. For example, in underwater acoustic communication, underwater waves use sound waves instead of radio waves because of the serious attenuation of underwater waves. Therefore, in underwater communication, the transmission of text, voice, image and other information needs to be converted into an electrical signal and then converted into an acoustic signal. At this time, in order to separate the speech signal in the water from noise such as waves, fish, and ships, the influence of the front and back states on the output should be considered. During the collection and research of marine sounds, the sound of fish as a signal for communication between fish schools should also consider the impact of the entire sequence on the output of the network. To this end, Bi-LSTM can be used to make full use of the context information in the sample for training.
\begin{figure}[!t]
	\centering
	\includegraphics[width=3.5 in]{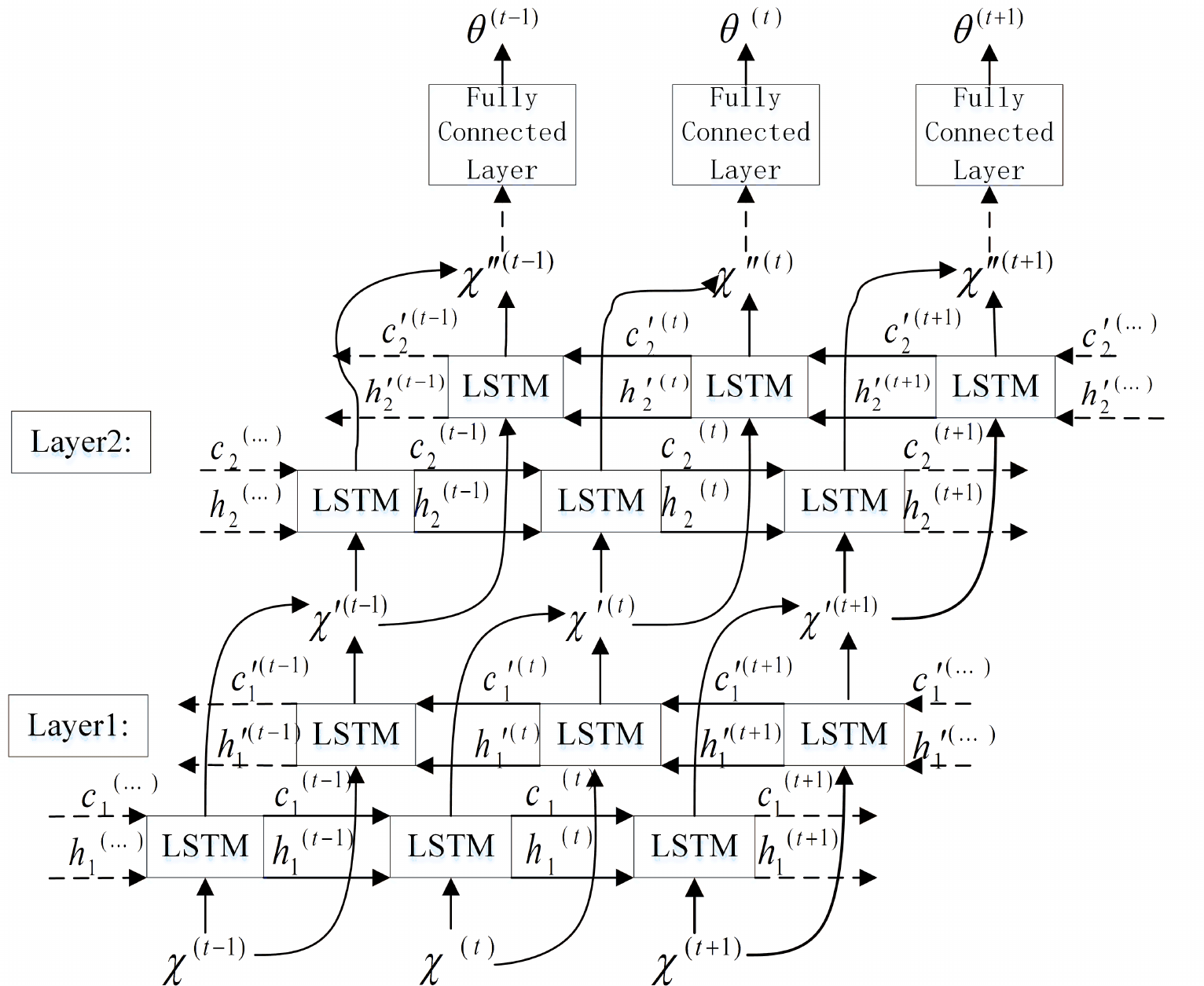}
	\caption{Bi-LSTM underwater acoustic separation network model diagram}
	\label{Bi-LSTM}
\end{figure}

The Bi-LSTM consists of two LSTMs of the same size and opposite start of the time series. Fig.\ref{Bi-LSTM} shows the structure of a water-acoustic separation network based on Bi-LSTM. Where $h^{(t)}$ represents the state of the sub-LSTM that propagates information from $t=1$ to $T$ (to the right) by time. ${h}'^{(t)}$ represents the state of the sub-LSTM in which the information moves backward from $t=T$ to 1 (to the left), and can be obtained by substituting the reverse sequence into equation \eqref{LSTM1}-\eqref{LSTM5}.

The specific operation of the unidirectional sub-LSTM layer is as follows. Given an input sequence $\mathbf{X}=\{X_{1},\ldots,X_{T}\}$, this model can be iteratively computed from $t=1$ to $T$, which is composed of the following:
\begin{equation}\label{LSTM1}
i_t=\sigma(W_{Xi}X_t+W_{hi}h_{t-1}+W_{ci}c_{t-1}+b_i),
\end{equation}
\begin{equation}
f_t=\sigma(W_{Xf}X_t+W_{hf}h_{t-1}+W_{cf}c_{t-1}+b_f),
\end{equation}
\begin{equation}
c_t=f_tc_{t-1}+i_ttanh(W_{Xc}X_t+W_{hc}h_{t-1}+b_c),
\end{equation}
\begin{equation}
o_t=\sigma(W_{Xo}X_t+W_{ho}h_{t-1}+W_{co}c_t+b_o),
\end{equation}
\begin{equation}\label{LSTM5}
h_t=o_ttanh(c_t),
\end{equation}
$W$ and $b$ are weights and biases, and $i$,$f$,$o$ and $c$ are the input gate, forget gate, output gate and cell activation vector respectively. $\sigma$ is the logistic sigmoid function.

Therefore, at each time point t, the output unit can obtain information about the past sequence with respect to the input ${h}'^{(t)}$ and the relevant information about the future sequence of the input $h^{(t)}$. After two sub-LSTM layers, we use a dense layer to obtain $\Theta_{t}$ which is the output of the $X_{i}$ as
\begin{equation}
  \Theta_{t}=\phi(W_hh_{t}^{l}+b_{\Theta}),
\end{equation}
where $h_{t}^{l}$ is the output of the final LSTM layer and $\phi$ is the Relu activation function. By minimizing the loss value, some parameters will adaptively change as the advancement of the learning process.

In the following experiments, the paper extracted the characteristics of the underwater acoustic signal using the above three networks (RNN, LSTM, Bi-LSTM) in the offline training phase. In the online test phase, combined with STFT and binary time-frequency masking methods, we obtained the corresponding experimental data of the three networks. Experiments have shown that the Bi-LSTM structure has the best performance.

\subsection{Online Test}

Different models are trained and applied to the traditional binary T-F masking framework. The processing flow of the method is basically the same as the processing flow of the binary T-F masking method. The main steps are as follows:

(1) Select the underwater acoustic signal in the test set for mixing to obtain a mixed underwater acoustic signal. The signal is de-equalized, normalized, and the signal is subjected to STFT (the parameters of the STFT in the test phase are consistent with the STFT parameters in the training phase), and finally $|X|$ is obtained as an input;

(2) Using the trained model, the original feature X of the signal is transformed into a new embedded feature $\Theta$. Since the new feature is just a matrix of dimension $T*FK$ when it is output from the network, in the actual processing, we need to reshape the data, convert its dimensions into $TF*K$, and facilitate subsequent cluster analysis;

(3) Cluster analysis. Clustering analysis of feature $\Theta$ using K-means algorithm;

(4) T-F masking. According to the set $\Omega_k$ obtained by clustering, the corresponding binary T-F masking matrix $M_k(t,f)$ is set and substituted into the equation \eqref{yk(t,f)}. Obtaining a T-F domain estimate of the source signal;

(5) Time domain recovery. The source signal $\tilde{S_k}(t,f)$ estimated in the above step is subjected to ISTFT estimation according to the equation \eqref{yk(t)} to obtain a time domain waveform $\tilde{s_k}(t)$ of the source signal.

The clustering algorithm is used to classify this feature of the neural network output so that the vector $\theta$ belonging to the same underwater sound source can be divided into a group. Set each vector of "similar" to 1, and set other vectors that are not similar to 0. The new array dimension is reconstructed into a matrix of $T*F$, which is the binary masking matrix corresponding to the water source.

\section{Experiments}
\label{sec:guidelines}

\subsection{Experimental conditions}

The experiment selected a hydroacoustic audio data in SHIPEAR as a data sample\cite{Santos2016ShipsEar}. Since its establishment, the database has been used for research on ship noise reduction, detection, identification, etc., especially for the application of deep learning technology \cite{shen2018compression}\cite{shen2018auditory}\cite{ke2018underwater}. The hydroacoustic data on this database was collected by a hydrologist from the Atlantic coast of northwestern Spain, by researcher David and others from the University of Vigo in Spain. The composition of the database is shown in table \ref{table_database}. The sonar audio, ship radiation noise and background noise form A, B, and C signals, respectively, and each audio is selected to be about 6 seconds in length for testing. The sample sampling rate is unified to 44100 Hz. In addition, we also simulated the binary time-frequency masking method. By comparing the effects of binary separation and multiple separation, the superior performance of the proposed method is proved. For the binary time-frequency masking method, this section selects three LFM signals for simulation, which facilitates the aliasing operation of the signals at time and frequency. Simulate three LFM signals, the sampling frequency is 50 kHz, and the time length is 1 s. The specific parameters are shown in table \ref{table_LFM}.

\begin{table}[!t]
	\caption{The composition of the database}
	\label{table_database}
	\centering
	\begin{tabular}{|c|c|}
		\hline
		Category             & Details                                \\ \hline
		Number of recordings & 90 segments                            \\ \hline
		Recording length     & 15 s to 10 minutes                     \\ \hline
		Number of ships      & 11                                     \\ \hline
		Background noise     & different depths and channel distances \\ \hline
	\end{tabular}
\end{table}

\begin{table}[!t]
	\caption{LFM signal parameters for binary time-frequency masking method simulation}
	\label{table_LFM}
	\centering
	\begin{tabular}{|c|c|c|c|c|c|}
		\hline
		Signals  & Frequency Range & Launch time & duration \\ \hline
		LFM 1    & 6 - 8 kHz       & 0.1 s       & 0.3 s    \\ \hline
		LFM 2    & 6.5 - 10 kHz    & 0.5 s       & 0.2 s    \\ \hline
		LFM 3    & 12 - 15 kHz     & 0.6 s       & 0.3 s    \\ \hline
	\end{tabular}
\end{table}

In the training stage, we try to train the model with the maximum mixture number of 3. So, we randomly select two or three files from the training set to mix in every iteration. Then we use the model to separate every possible underwater acoustic mixing source. We design the network structure with two LSTM layers with 600 hidden cells and a full connection layers with 100 cells corresponding with the embedding dimension $K$. Stochastic gradient descents with momentum 0.9 and fixed learning rate $10^{-5}$ was used for training. The Relu function is used as the activation function for the output layer. n order to prevent the network from over-fitting and improve the generalization ability of the model, the input layer and the hidden layer's dropout parameters are set to 0.2 and 0.5 respectively. And add L2 regularization to the network, the parameter is set to $10^{-6}$. The training iterations of the model is 30.

In the test stage, the input feature $X$ is the log magnitude spectrum of the mixed underwater acoustic signal using STFT with 32ms frame length, 8 ms window shift, and the square root of the hanning window. Moreover, the mixture is separated into 100 frames with half overlap to ensure the local accuracy of output feature $\Theta$. The masks were obtained by clustering the row vectors of the feature $\Theta$. The number of clusters is set to the number of sources in the mixture.

\subsection{Metircs}
To evaluate the quality of the source separation, we use three quantitative criteria: 1) the Preserved-Signal Ratio (PSR$\in[0,1]$), representing the quality of the mask preserving the target source. 2)the Signal-to-Interference Ratio (SIR$\in[0,\infty)$), representing the quality of the mask suppressing the interfering sources. 3)The similarity coefficient $\xi$ estimates the similarity between the signal $y_i(t)$ and the source signal $x_j(t)$

\textbf{PSR:} The Preserved-Signal Ratio (PSR) is used to measure the degree of protection of the masking matrix $M_k$ to the target signal $X_k(t,f)$. The mathematical equation is expressed as follows:
\begin{equation}\label{PSR}
PSR=\frac{\|M_k(t,f)X_k(t,f)\|^2}{\|X_k(t,f)\|^2}
\end{equation}
The PSR characterizes the amount of energy remaining after the target signal passes through the masking matrix. In the equation, $\left \| \cdot  \right \|^2$ represents a double integral operation, that is, $\left \| f(x,y) \right \|^2=\iint \left | f(x,y) \right |^2dxdy$. PSR satisfies $0\leq PSR\leq 1$. If the estimated masking matrix $M_k$ satisfies the relationship $\hat{M_k} \subseteq  M_k$ with the actual masking matrix $\hat{M_k}$ , then PSR=1.

\textbf{SIR:} The SIR indicates the suppression of the interference source by the masking matrix. An interference source composed of other source signals other than the source signal $x_k(t)$ is denoted by $v_k(t)$, and the corresponding T-F domain is expressed as $V_k(x,y)$ . The signal-to-interference ratio for the masking matrix M is defined as follows:
\begin{equation}\label{SIR}
SIR_M=\frac{\|M_k(t,f)X_k(t,f)\|^2}{\|M_k(t,f)V_k(t,f)\|^2}
\end{equation}
$SIR_M$ is a value greater than or equal to 0. The larger the value, the better the separation performance. When the masking matrix is completely suppressed against other source signals, $SIR_M=\infty $. In the T-F mask separation method, good separation performance requires that the T-F information of the source signal be preserved as much as possible, and the interference source can be suppressed, that is, the PSR is close to 1, and the $SIR_M$ is as large as possible.

\textbf{$\xi$:} Similarity coefficient $\xi$
\begin{equation}
\xi _{ij}=\xi(y_i,x_j)=\frac{\left | \sum_{t=1}^{n}y_i(t)x_j(t) \right |}{\sqrt{\sum_{t=1}^{n}y_{i}^{2}(t)\sum_{t=1}^{m}x_{j}^{2}(t)}}
\end{equation}

If $\xi_{ij}=1$, it means that the i-th estimated signal is exactly the same as j source signals. If $\xi_{ij}=0$, it means that $y_i(t)$ and $s_j(t)$ are completely inconsistent. In the actual situation, due to the existence of the estimated difference, the separation performance of the similarity coefficient is generally close to 1, and the worst is 0. Generally, these coefficients constitute a similarity coefficient matrix. If only one similarity coefficient of each row in the matrix tends to 1, and the other tends to 0, the separation performance is good.

\subsection{Results}

\subsubsection{Binary signal separation using binary time-frequency masking method}

\begin{figure}[!t]
	\centering
	\subfloat[]{\includegraphics[width=1.5in]{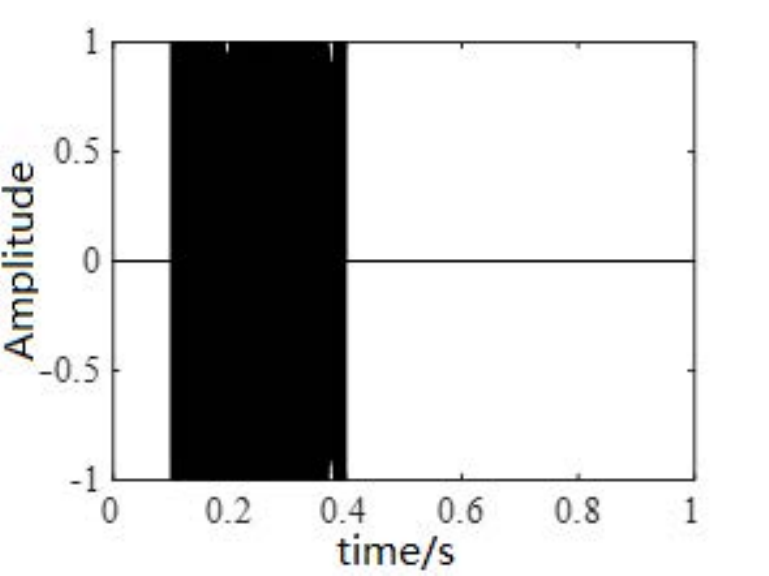}%
		\label{LFM1t}}
	\hfil
	\subfloat[]{\includegraphics[width=1.5in]{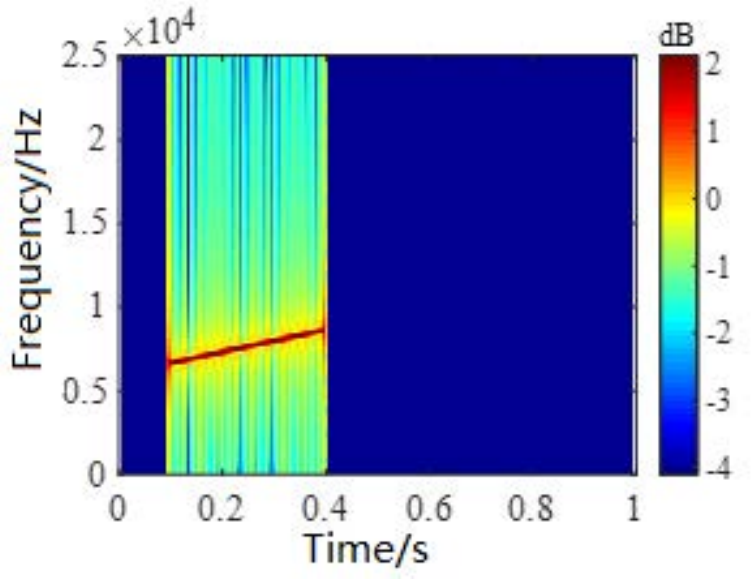}%
		\label{LFM1tf}}
	\hfil
	\subfloat[]{\includegraphics[width=1.5 in]{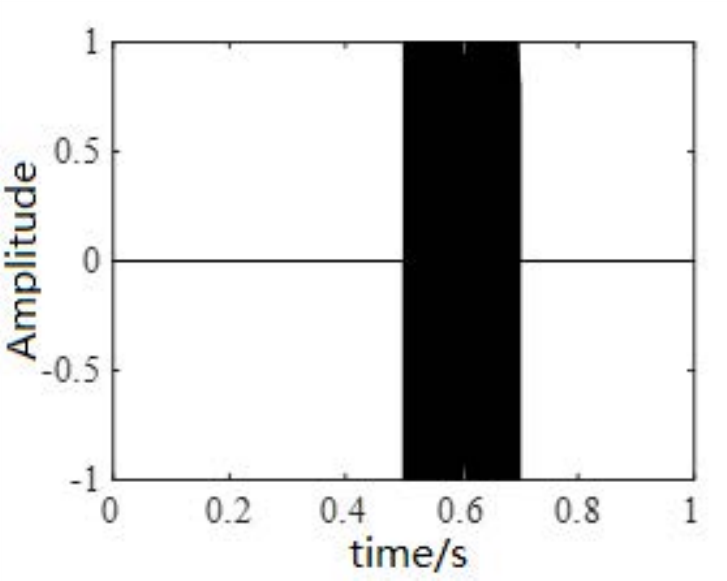}%
		\label{LFM2t}}
	\hfil
	\subfloat[]{\includegraphics[width=1.5 in]{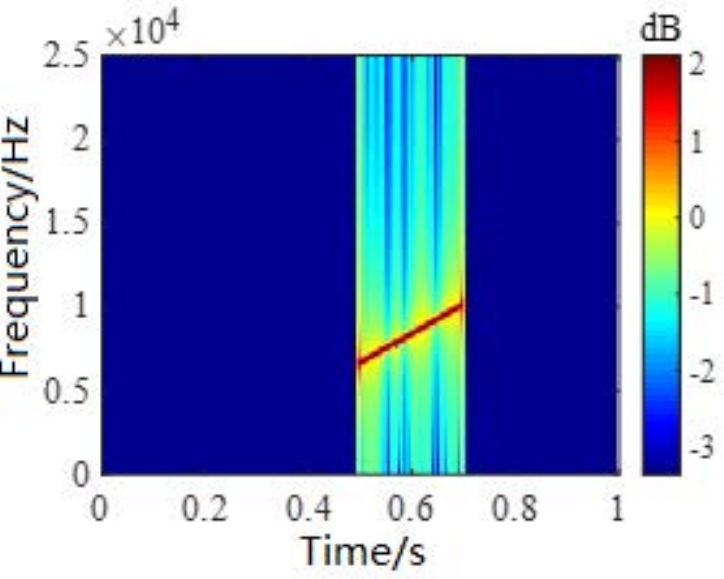}%
		\label{LFM2tf}}
	\hfil
	\subfloat[]{\includegraphics[width=1.5 in]{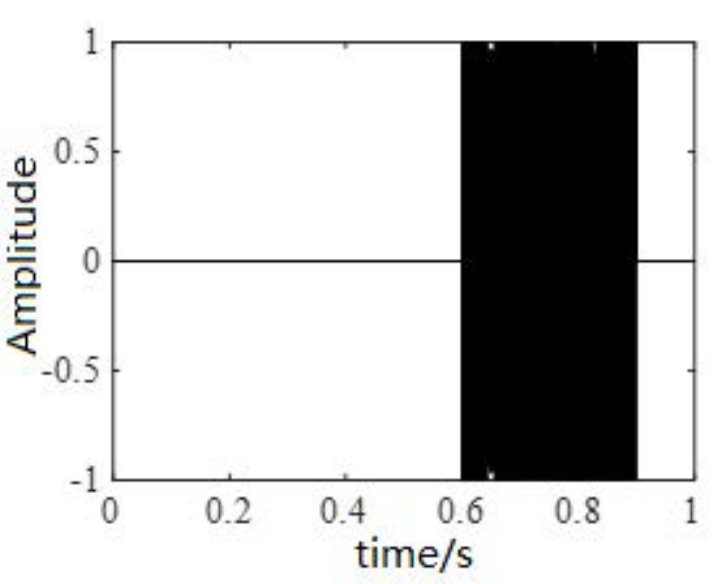}%
		\label{LFM3t}}
	\hfil
	\subfloat[]{\includegraphics[width=1.5 in]{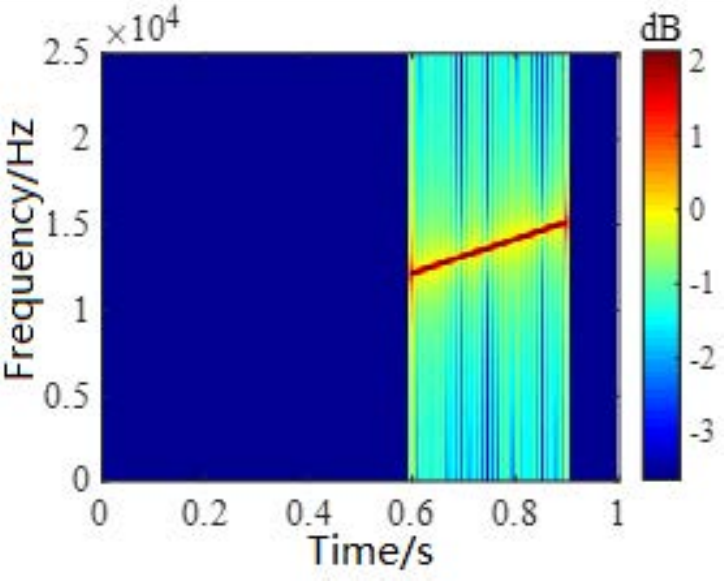}%
		\label{LFM3tf}}
	\caption{Time domain waveform and time-frequency diagrams of source signals}
	\label{LFM}
\end{figure}

The experiments first simulated three LFM signals that satisfy the energy-dominated condition. The time-domain waveform and time-frequency diagrams of the simulation signals are shown in Fig.\ref{LFM}. Fig.\ref{LFM}(a)(c)(e) are time-domain waveforms of the three signals, and Figures.\ref{LFM}(b)(d)(f) are time-frequency diagrams of the three signals.


\begin{figure}[!t]
	\centering
	\subfloat[]{\includegraphics[width=1.5 in]{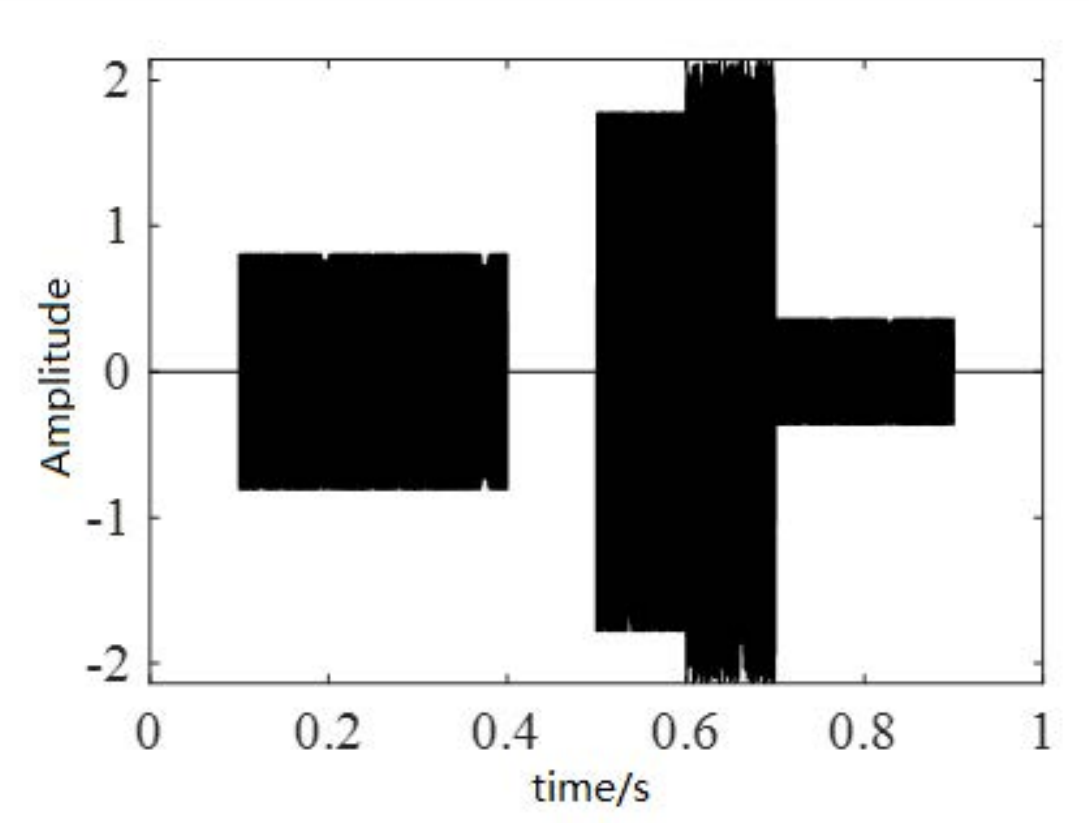}%
		\label{2-9a}}
	\hfil
	\subfloat[]{\includegraphics[width=1.5 in]{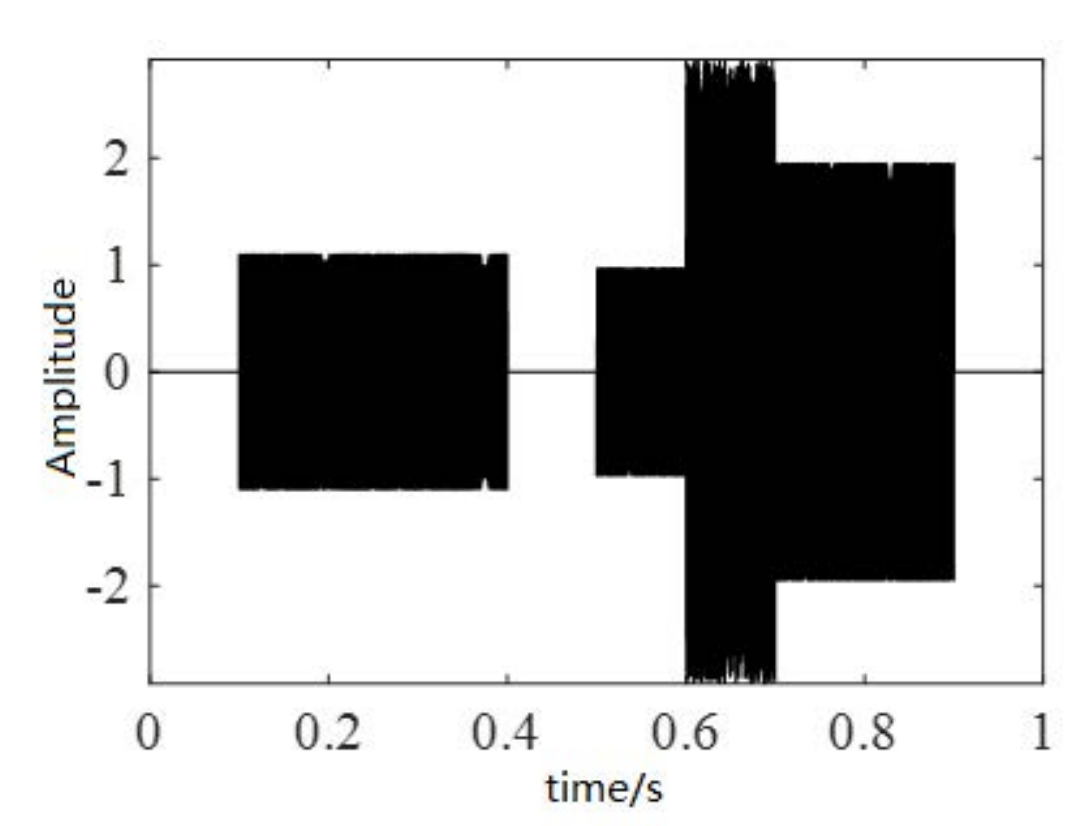}%
		\label{2-9c}}
	\hfil
	\subfloat[]{\includegraphics[width=1.5 in]{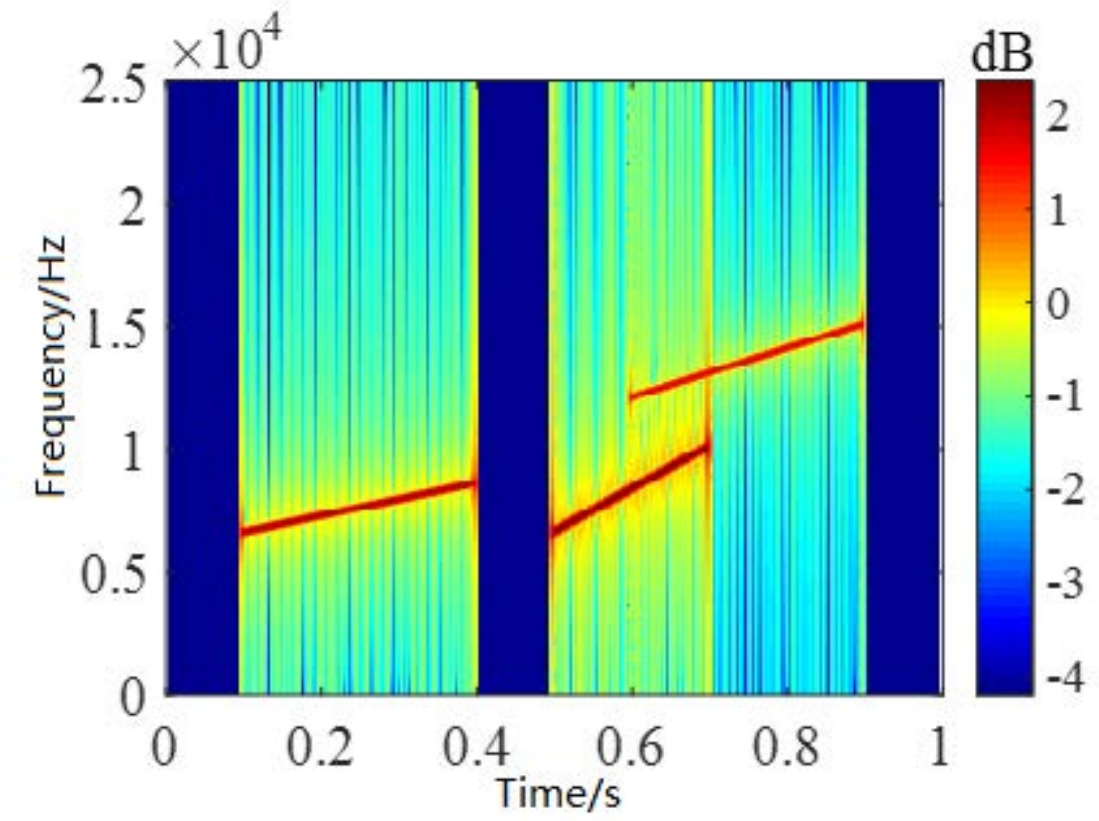}%
		\label{2-9b}}
	\hfil
	\subfloat[]{\includegraphics[width=1.5 in]{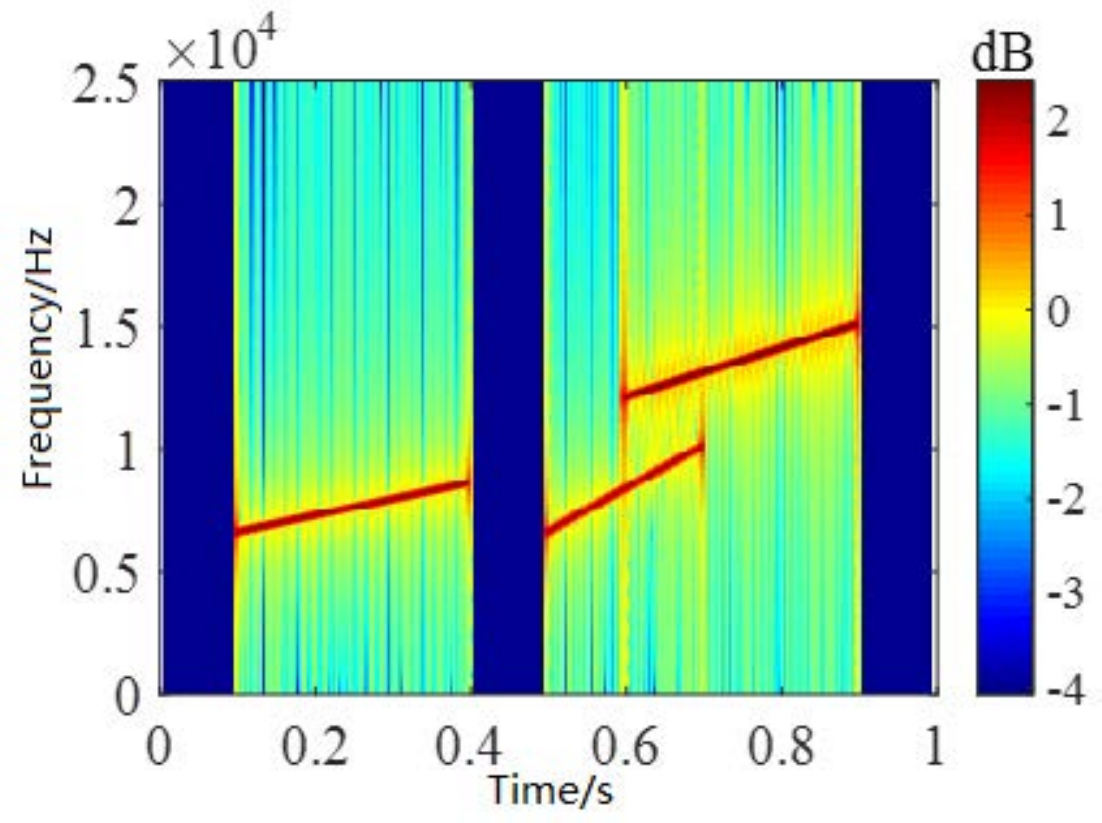}%
		\label{2-9d}}
	\hfil
	\caption{Time-domain waveforms and time-frequency diagrams of observed signals}
	\label{figure2-9}
\end{figure}

\begin{figure}[!t]
	\centering
	\subfloat[]{\includegraphics[width=1.5 in]{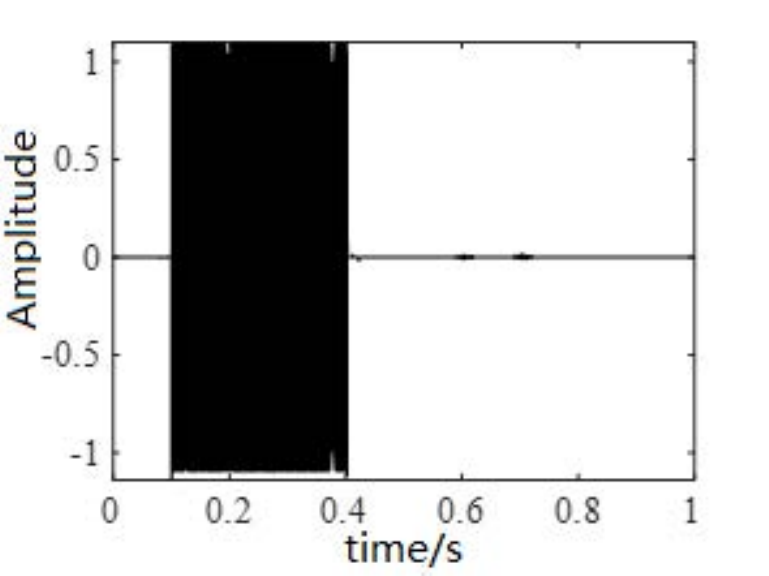}%
		\label{2-10a}}
	\hfil
	\subfloat[]{\includegraphics[width=1.5 in]{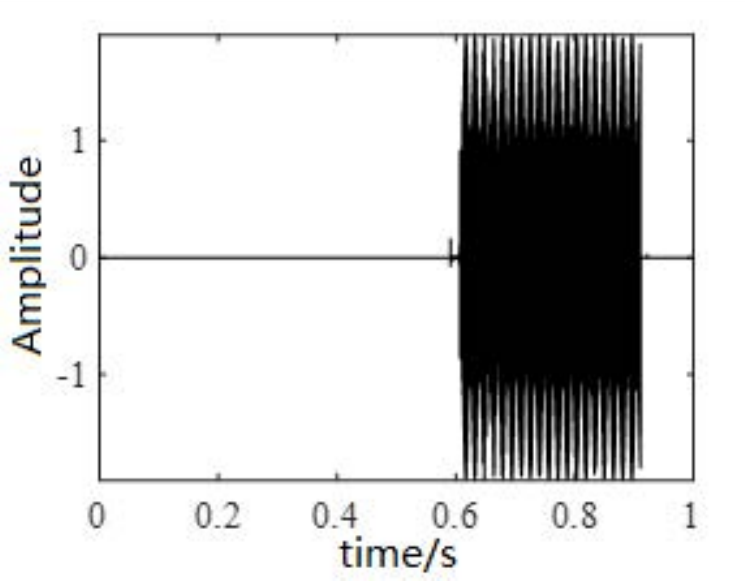}%
		\label{2-10c}}
	\hfil
	\subfloat[]{\includegraphics[width=1.5 in]{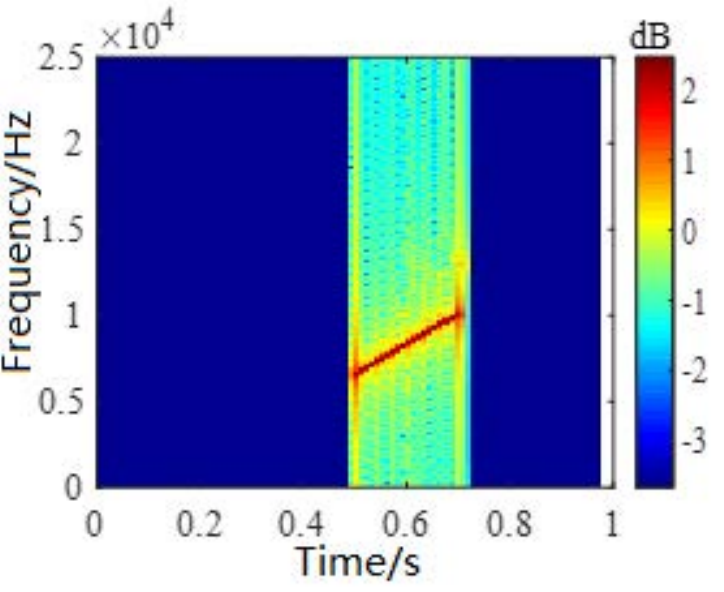}%
		\label{2-10e}}
	\hfil
	\subfloat[]{\includegraphics[width=1.5 in]{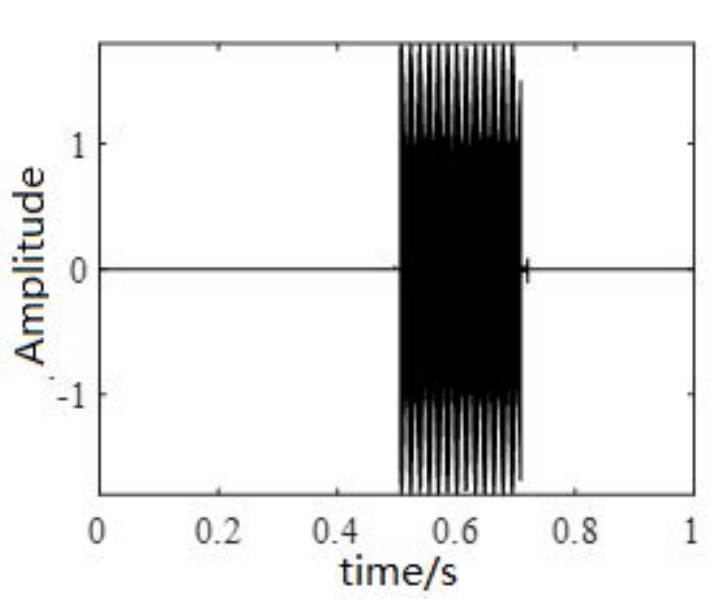}%
		\label{2-10b}}
	\hfil
	\subfloat[]{\includegraphics[width=1.5 in]{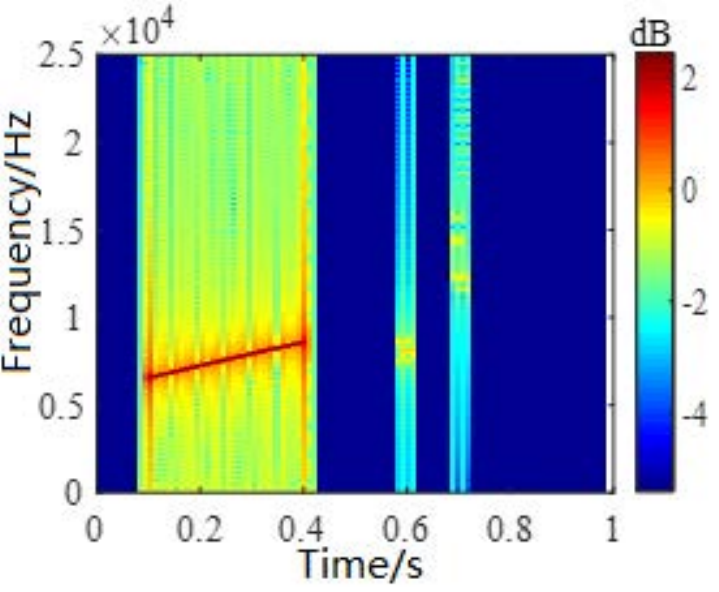}%
		\label{2-10d}}
	\hfil
	\subfloat[]{\includegraphics[width=1.5 in]{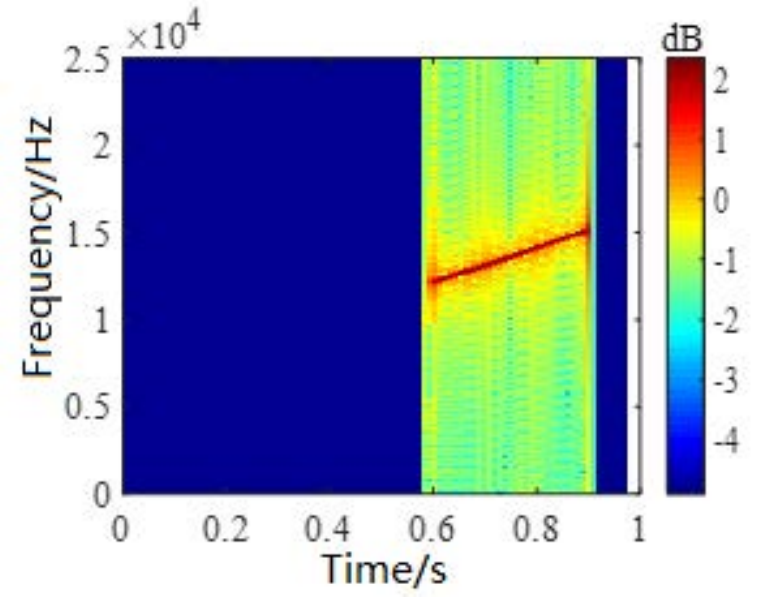}%
		\label{2-10f}}
	\hfil
	\caption{Time-domain waveforms and time-frequency diagrams of estimated signals}
	\label{figure2-10}
\end{figure}

The randomly generated mixing matrix is linearly mixed according to equation \ref{mix2-2}. The time-domain waveform and time-frequency diagrams of the observed signals are shown in Fig.\ref{figure2-9}. Select the hamming window for the observation signal, perform 512-point STFT transformation, and set 25\% overlap to obtain the time-frequency characteristics. According to the equations \ref{2-43} and \ref{2-45}, take the magnitude of the observation signal and the phase difference to form a feature vector, and finally obtain an estimate signal. The time-domain waveforms and time-frequency diagrams of the observed and estimated signals are shown in Fig.\ref{figure2-9}. It can be seen from the results that when the signal meets the sparsity condition, the binary signal can be recovered by using a binary time-frequency masking algorithm. The time-domain waveform and time-frequency diagram of the estimated signal are shown in Fig.\ref{figure2-10}.

\begin{equation}
	x(t)=As(t)+n(t)
	\label{mix2-2}
\end{equation}

From the effect diagram of the estimated signal, the source signal can be basically recovered using the binary time-frequency masking method. The No. 2 source signal is aliased with the No. 1 source signal and No. 3 source signal in the frequency domain and time domain, respectively, so the information will be affected somewhat, but it can basically be recovered from the mixed signal.

Correlation coefficients $\xi$, PSR, and SIRM were measured under different signal-to-noise ratios. The results are shown in table \ref{sepSNR}. It can be seen that when there is no noise, each signal can be well recovered. The two parameters of PSR and SIR indicate that the method can correctly divide the time-frequency region of each signal, that is, the obtained masking matrix accurately covers the time of the signal. Frequency information. Once noise is added, the performance deteriorates, and the PSR reduction is small, but the SIRM reduction is the most obvious. It means that after adding noise, the estimated masking matrix not only loses some information of the signal itself, but also receives time-frequency information of other signals.

\begin{table}[!t]
	\caption{Separation performance at different SNR}
	\label{sepSNR}
	\centering
	\begin{tabular}{|c|c|c|c|c|c|c|}
		\hline
		SNR/dB & 0    & 5         & 10         & 15    & 20     & No noise \\ \hline
		$\xi$  & 0.60 & 0.62      & 0.74       & 0.81  & 0.89   & 0.98     \\ \hline
		$PSR$    & 0.71 & 0.72      & 0.82  & 0.85  & 0.90   & 0.98     \\ \hline
		$SIR_{M}$    & 5.82 & 5.47      & 15.56 & 27.93 & 316.21 & 24193.72 \\ \hline
	\end{tabular}
\end{table}

\subsubsection{Binary and multivariate signal separation using the proposed method}
\begin{figure}[!t]
	\centering
	\subfloat[]{\includegraphics[width=1.5 in]{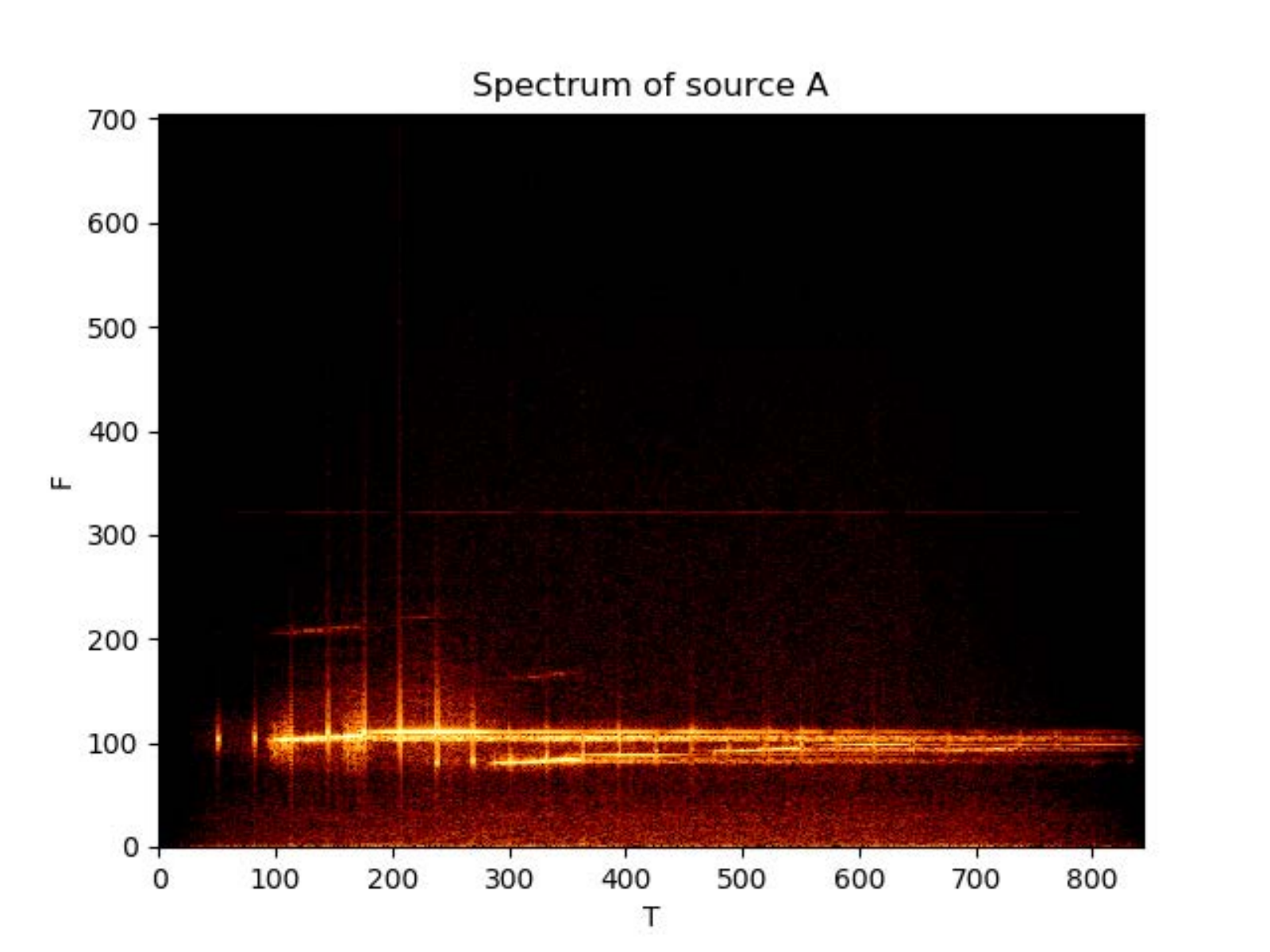}%
		\label{ACG_a}}
	\hfil
	\subfloat[]{\includegraphics[width=1.5 in]{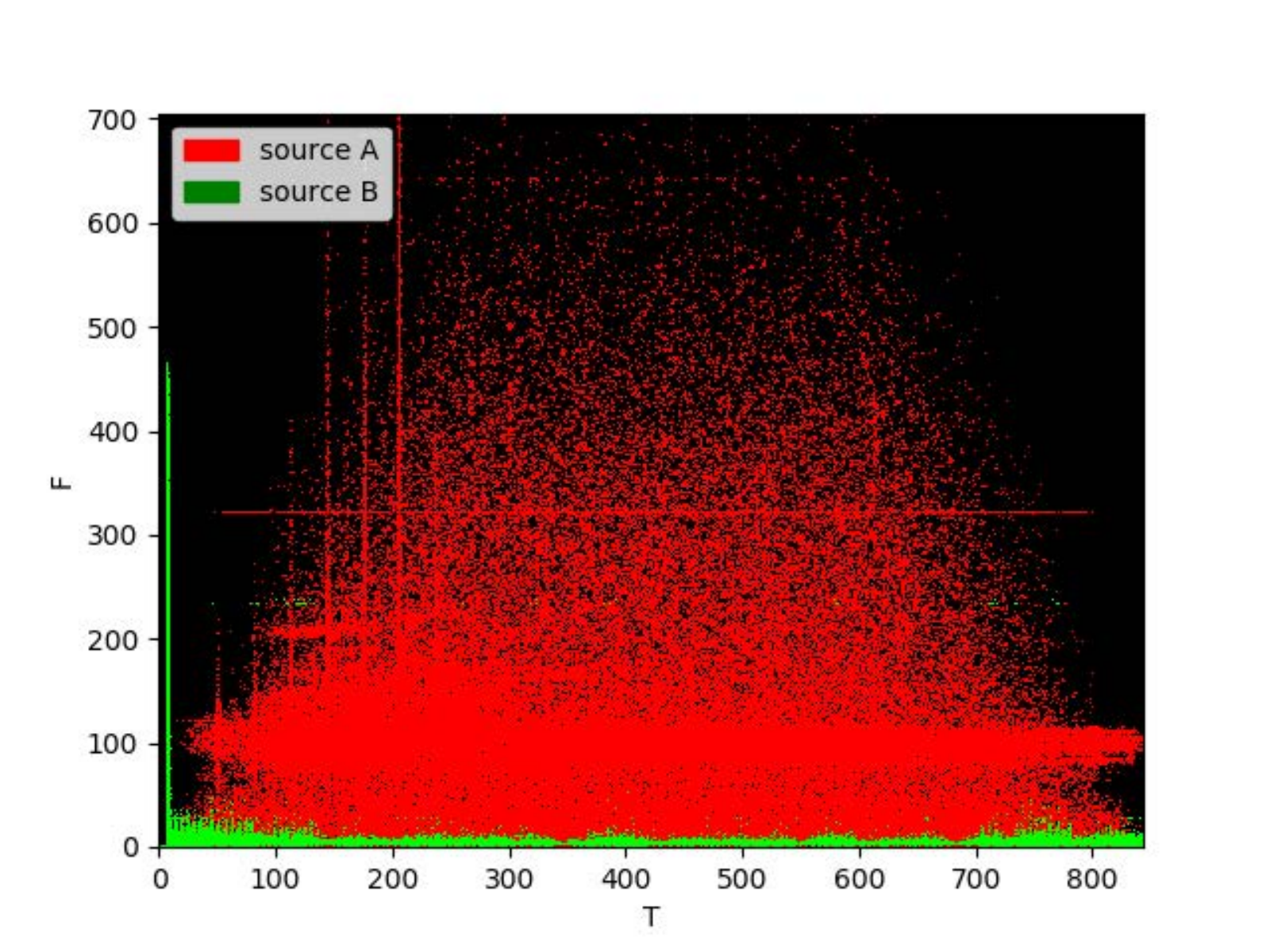}%
		\label{ACG_b}}
	\hfil
	\subfloat[]{\includegraphics[width=1.5 in]{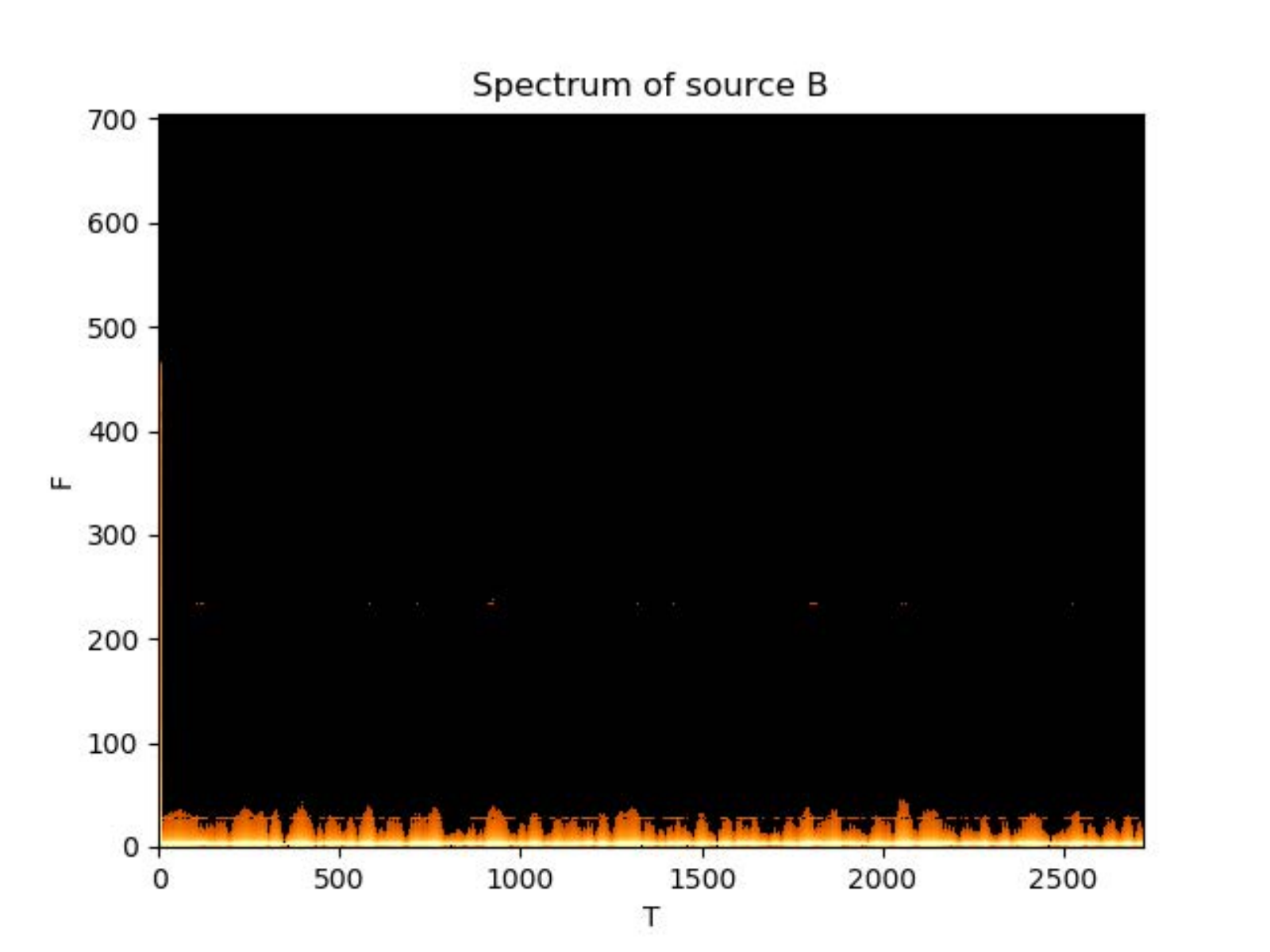}%
		\label{ACG_c}}
	\hfil
	\subfloat[]{\includegraphics[width=1.5 in]{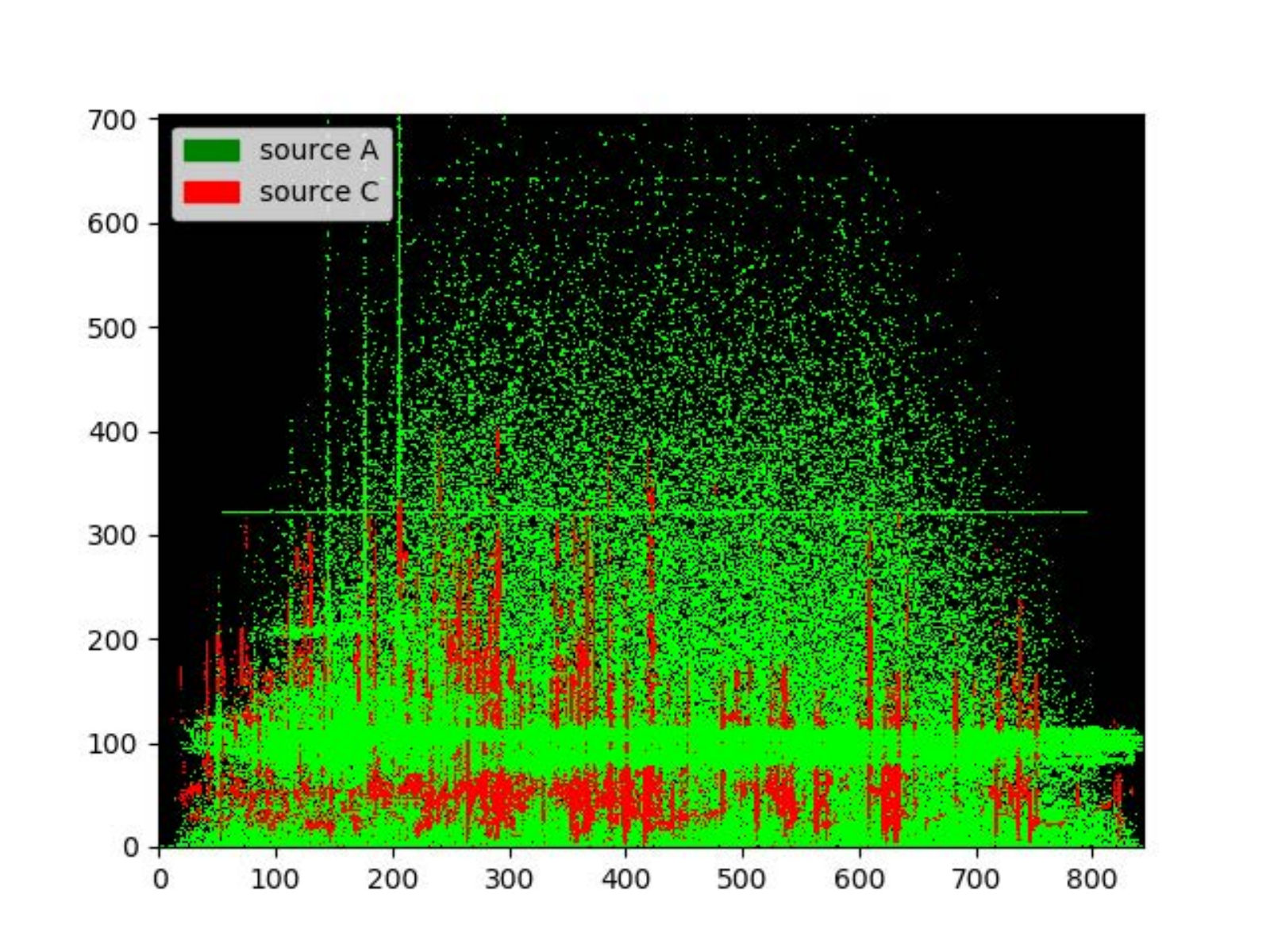}%
		\label{ACG_d}}
	\hfil
	\subfloat[]{\includegraphics[width=1.5 in]{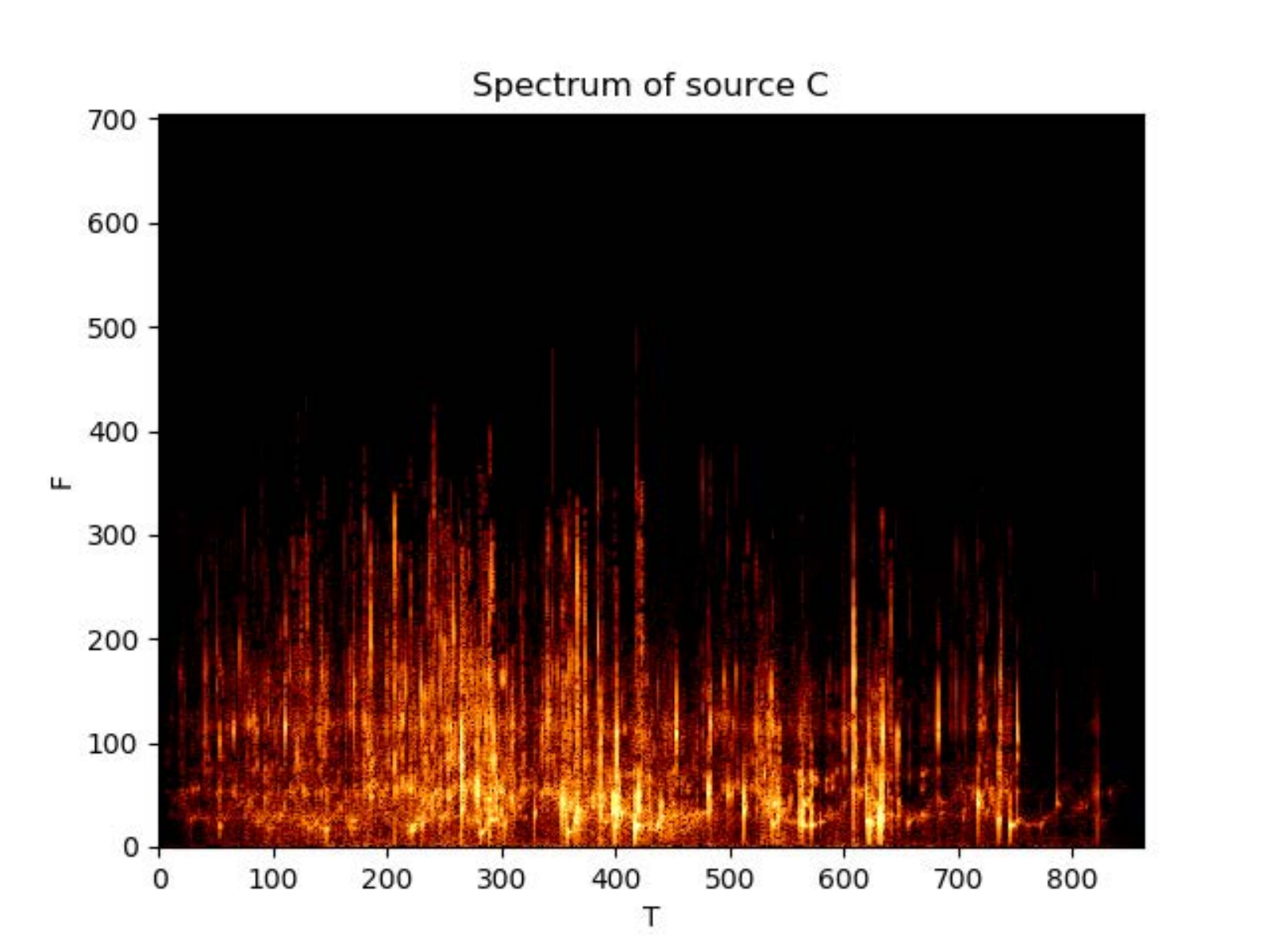}%
		\label{ACG_e}}
	\hfil
	\subfloat[]{\includegraphics[width=1.5 in]{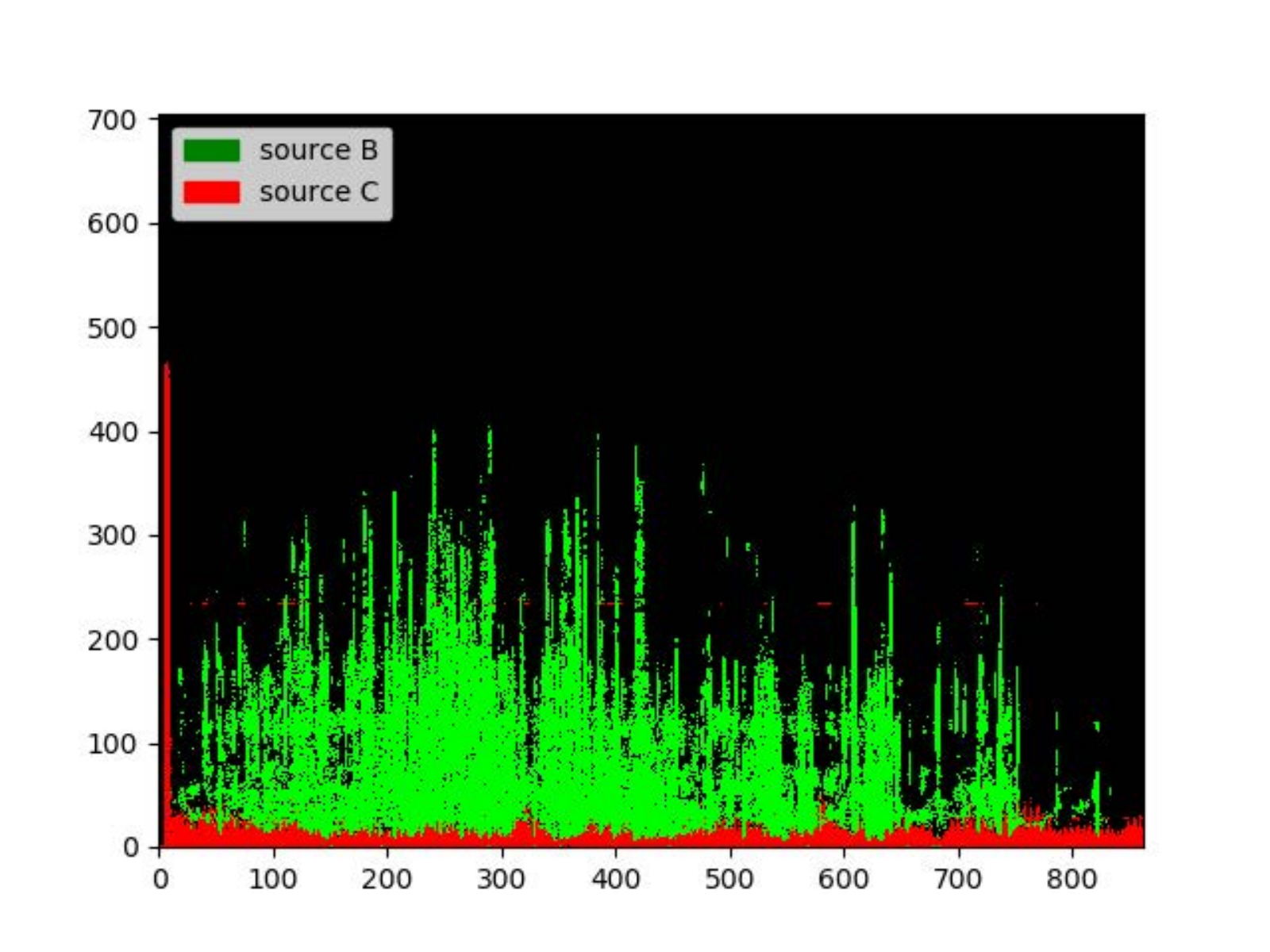}%
		\label{ACG_f}}
	\caption{Visualization of separating two-unknown-source mixtures}
	\label{ACG}
\end{figure}

Next, we separate mixing signals of two sources. The visualization result can be seen in Fig.\ref{ACG}. We list all possible combinations and observe the corresponding effect of separation. Fig.\ref{ACG}(a)(c)(e) are the spectrum of sources A,B,C separately. Fig.\ref{ACG}(b)(d)(f) print the separation results of those pairwise mixtures of A, B, C respectively. Compared with the original spectrums, every source can be perfectly separated.


In the table \ref{table_ex1and1}, we illustrate the demixing performance about separating two unknown sources using the metrics mentioned in equation \eqref{PSR} and \eqref{SIR}. It shows that our proposed method has better performance on separating two known sources, which indicates that this approach is different from many separation algorithms based on deep learning. At the same time, it can spread well to unknown sources separation without any specific adaptation methods. SIR is infinity because the interfering sources are suppressed sufficiently and make the denominator close to 0 according to the equation \eqref{SIR}.

\begin{figure}[!t]
	\centering
	\includegraphics[width=3 in]{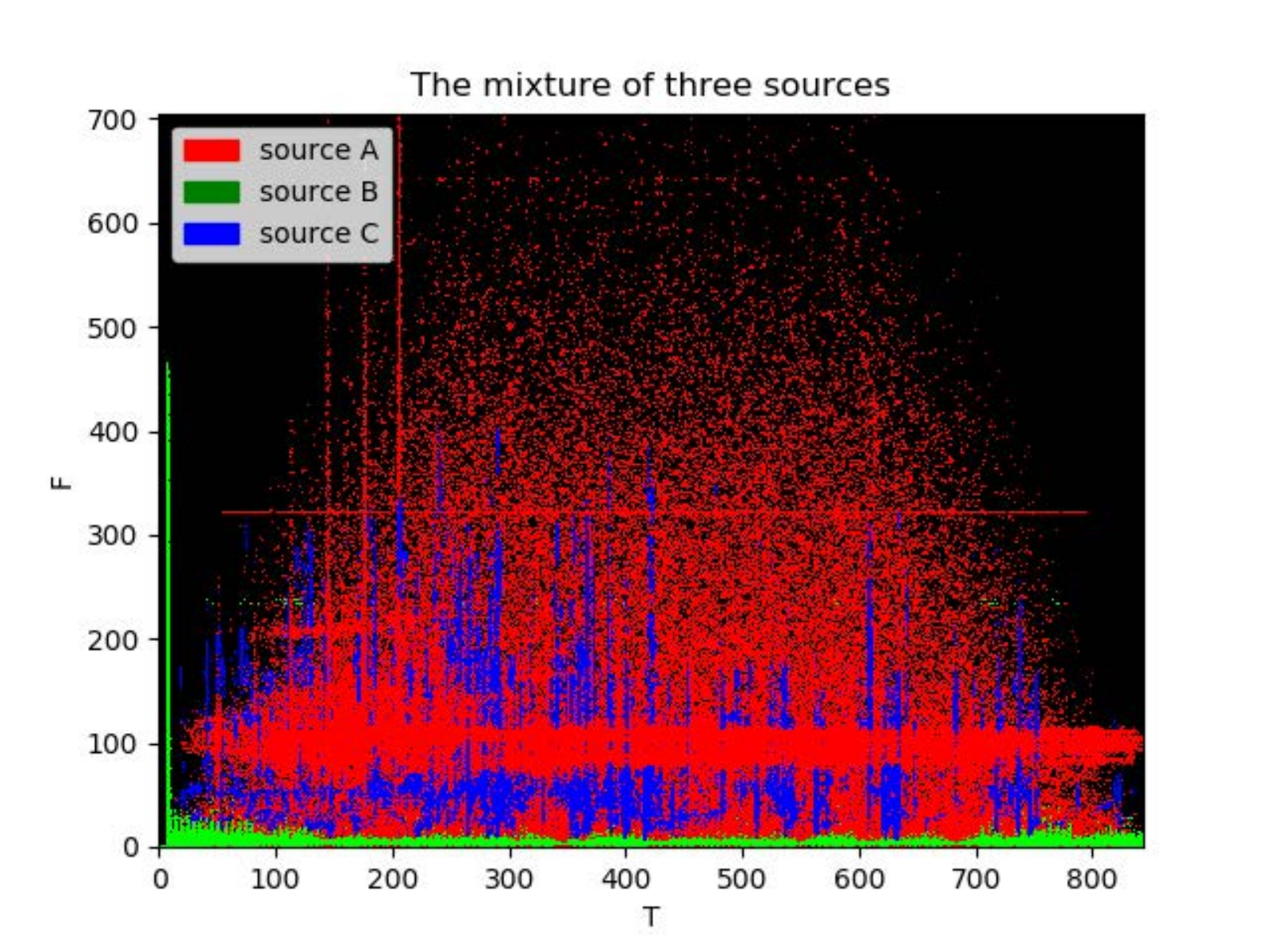}%
	\caption{Separation result of three-source mixtures}
	\label{ABC}
\end{figure}

\begin{figure}[!t]
	\centering
	\includegraphics[width=3 in]{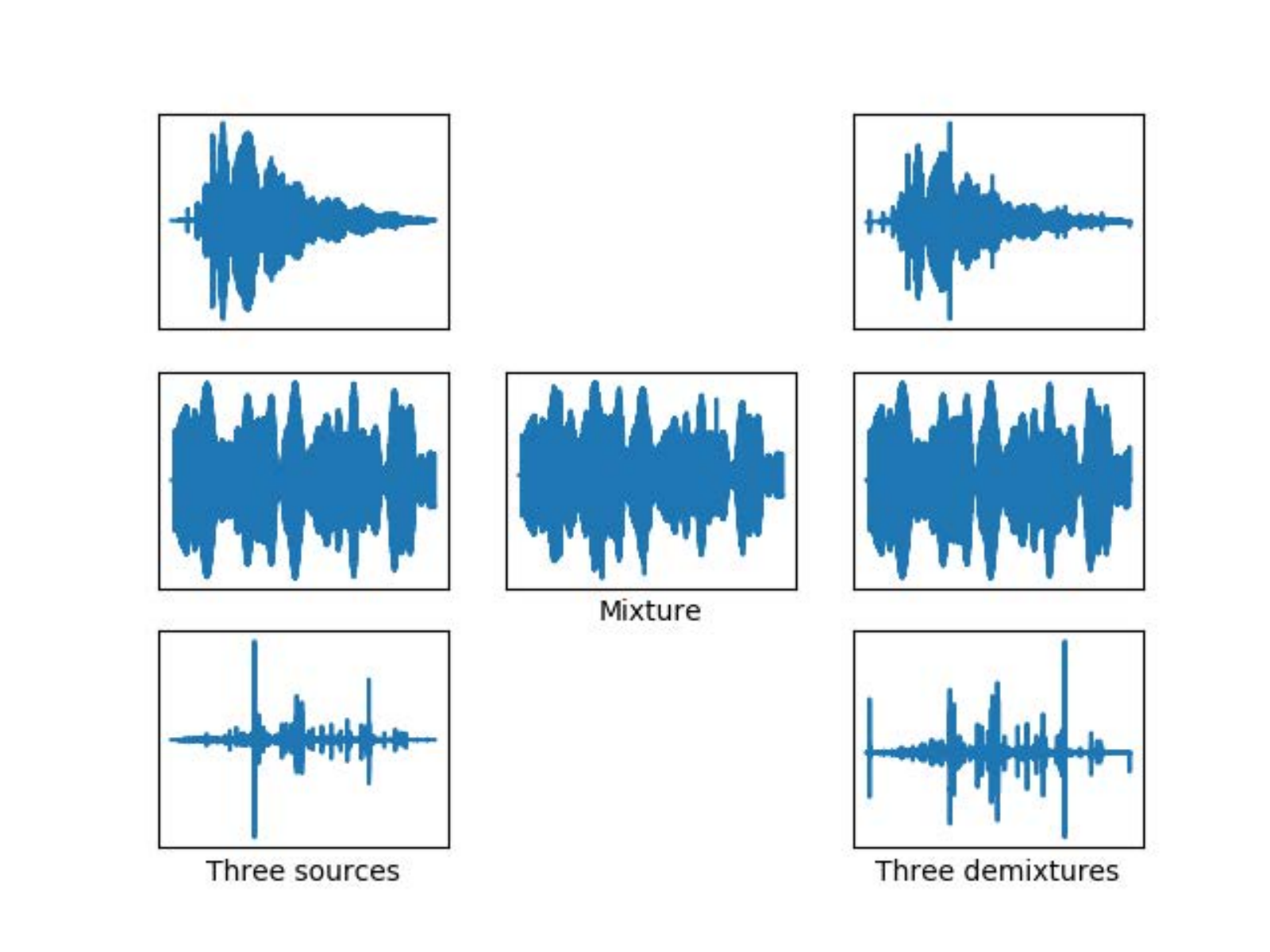}
	\caption{Three sources, mixture and three demixtures}
	\label{wave}
\end{figure}

Furthermore, we separate mixing signals of three sources. Fig.\ref{ABC} and Fig.\ref{wave} show the example of separating the mixtures of three sources. From the comparison of Fig.\ref{ACG} and Fig.\ref{ABC}, it can be seen that the time and frequency points of each source can be basically found. The overlap between the source signal C and the source signal A is relatively large in the time-frequency domain. However, signal A is dominated by energy at these overlapped time and frequency points, so it will not be disturbed by signals and can be basically recovered, but some information of signal C is lost. In fact, compared with background noise, people are more concerned about the loss of sonar echo signal. Therefore, it is permissible to sacrifice part of signal C in practical application. The overlap between signal B and signal A and C in frequency domain is the least, and the separation performance is the best. However,in order to prove that using deep learning method to separate underwater acoustic sources can achieve a breakthrough, we also show the results with the traditional binary T-F mask approach. In the table \ref{table_ex1and2}, the first one is our approach and the second is the traditional approach. It is clear that our proposed method outperforms the traditional method which even can't separate the source C and A very well. What's more, compared with table \ref{table_ex1and1}, when we separate more sources, the performance doesn't drop too much. So, the proposed model can scale up to more sources. Thus, it is appropriate for real world applications when the number of sources is not fixed.
\begin{table}[!t]
	
	\caption{The demixing performance in experiment }
	\label{table_ex1and1}
	\centering
	\renewcommand{\arraystretch}{1.4}
	\begin{tabular*}{\columnwidth}{l@{\extracolsep{\fill}}c@{\extracolsep{\fill}}c@{\extracolsep{\fill}}c@{\extracolsep{\fill}}c@{\extracolsep{\fill}}c}
		\hline
		\hline
		Sources&SIR in (dB)&SIR out(dB)& SIR gain (dB)& PSR\rule{0pt}{10pt}\\
		\hline
		Source A&-14.28&$\infty$&$\infty$&0.93&\rule{0pt}{10pt}\\
		Source B&14.28&$\infty$&$\infty$&0.92&\\
		Source C&-13.74&$\infty$&$\infty$&0.90&\\
		\hline
		\hline
	\end{tabular*}
\end{table}

\begin{table}[!t]
	
	\caption{Comparison of the Demixing Performance in Experiment 3 (top) With That of conventional T-F mask approach (bottom).}
	\label{table_ex1and2}
	\centering
	\renewcommand{\arraystretch}{1.4}
	\begin{tabular*}{\columnwidth}{l@{\extracolsep{\fill}}c@{\extracolsep{\fill}}c@{\extracolsep{\fill}}c@{\extracolsep{\fill}}c@{\extracolsep{\fill}}c}
		\hline
		\hline
		Sources&SIR in (dB)&SIR out(dB)& SIR gain (dB)& PSR\rule{0pt}{10pt}\\
		\hline
		Source A&-14.29&$\infty$&$\infty$&0.94&\rule{0pt}{10pt}\\
		Source B&14.10&$\infty$&$\infty$&0.93&\\
		Source C&-28.18&$\infty$&$\infty$&0.90&\\
		\hline
		\hline
		Source A&-14.29&13.93&28.22&0.81&\rule{0pt}{10pt}\\
		Source B&14.10&$\infty$&$\infty$&0.93&\\
		Source C&-28.18&2.42&30.6&0.29\\
		\hline
		\hline
	\end{tabular*}
\end{table}

\begin{figure}[!t]
	\centering
	\includegraphics[width=3.5 in]{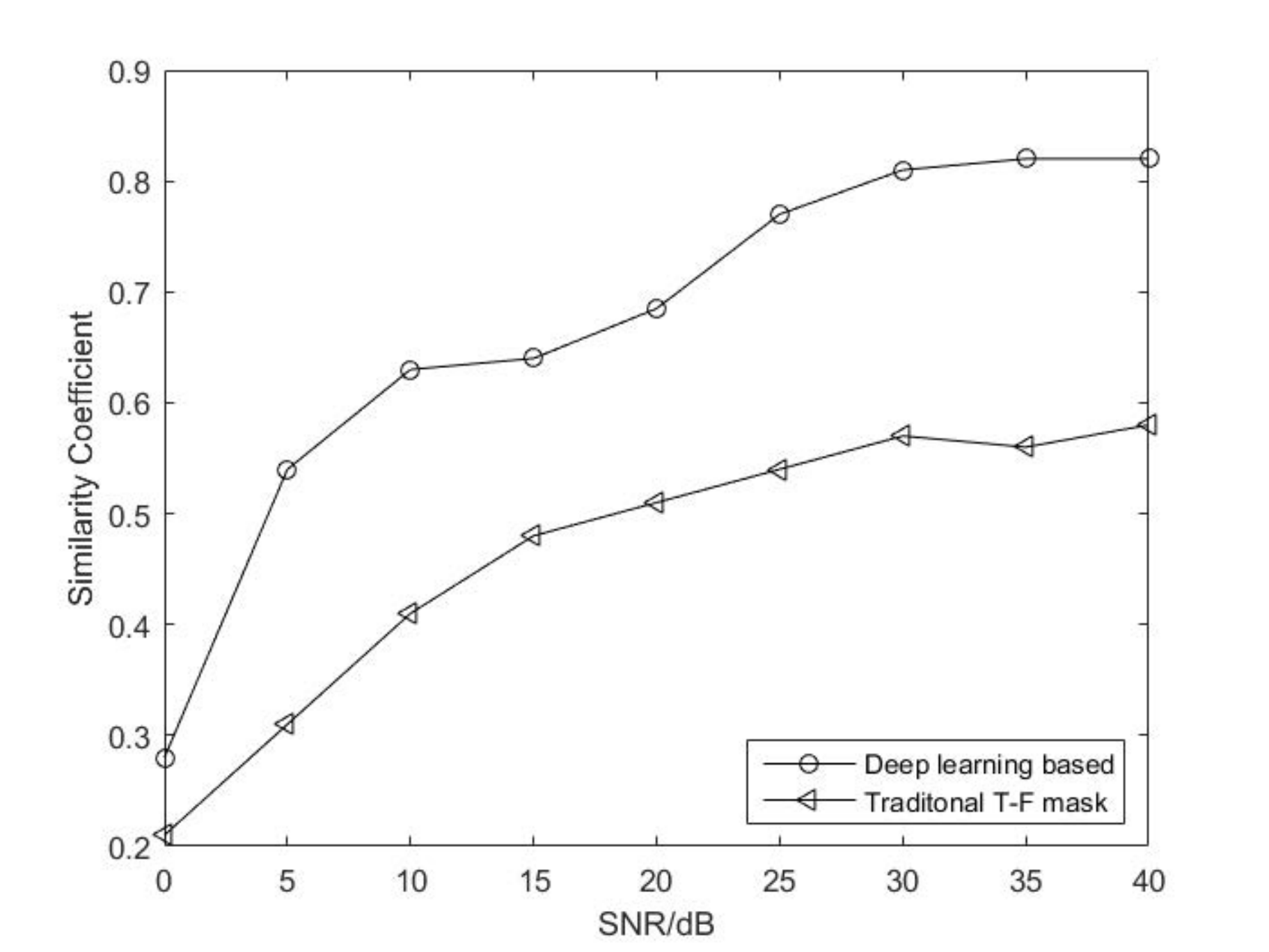}
	\caption{Comparison of similarity coefficients under different noise background}
	\label{similarity}
\end{figure}

In addition, considering that the mixed signal will be interfered by other unknown noises in the actual processing, gaussian noise signals of 0-40dB are added to the mixed signal to analyze the separation performance under different SNR conditions. Meanwhile, compared with the traditional binary T-F masking method, the similarity coefficient is used as the measurement standard, and the results are shown in Fig.\ref{similarity}. When the noise background is relatively strong, both the deep learning-based separation method and the traditional separation method will be greatly affected. As the noise is reduced, the estimated signal becomes clear gradually. Compared with the above separation situation, this test signal has larger aliasing in both frequency domain and time domain. So, the traditional time-frequency masking method has a poor separation performance, and its final average similarity coefficient is stable at about 0.6. The deep learning-based separation method can divide each target signal according to the energy dominant condition, so it has better separation performance on the whole.

\begin{figure}[!t]
	\centering
	\subfloat[]{\includegraphics[width=1.5 in]{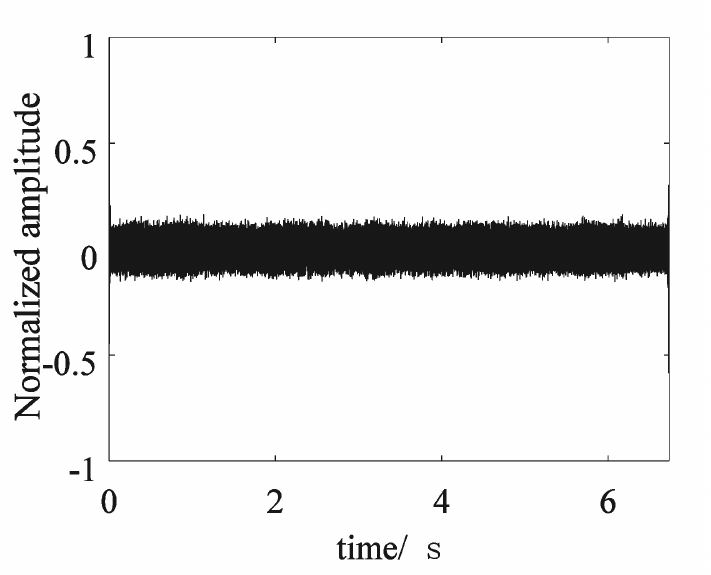}%
		\label{difk_a}}
	\hfil
	\subfloat[]{\includegraphics[width=1.5 in]{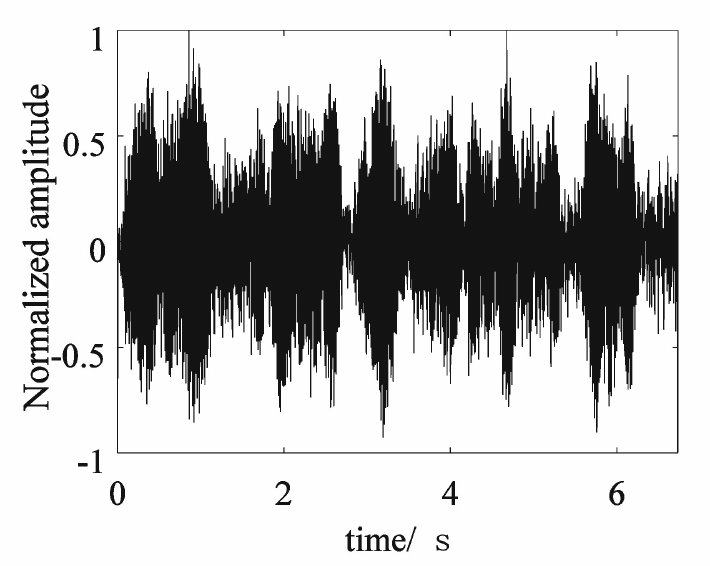}%
		\label{difk_b}}
	\hfil
	\caption{The separation results of signal C when different K values are set. (a)K=3; (b)K=4}
	\label{difk}
\end{figure}

Under the condition of unknown noise, the separated signal will still carry noise and affect the performance. It is found that the noise can be separated as long as the number of clusters is increased when the clustering algorithm is used. Take the case of adding gaussian white noise with SNR of 0dB as an example. When the signal is divided into three categories, each signal will carry noise. Among them, signal C suffers the largest interference and has a very weak energy, as shown in Fig.\ref{difk}(a). Increase the number of clusters to 4, that is, set the K value of K-means clustering algorithm to 4. The proposed method can also separate noise from three source signals, and recover the basic shape of signal C, as shown in Fig.\ref{difk}(b). It can be proved that the proposed method can not only perform well in the separation of multivariate signals, but also work effectively in the presence of certain noise interference.

\begin{figure}
	\centering
	\includegraphics[width=3 in]{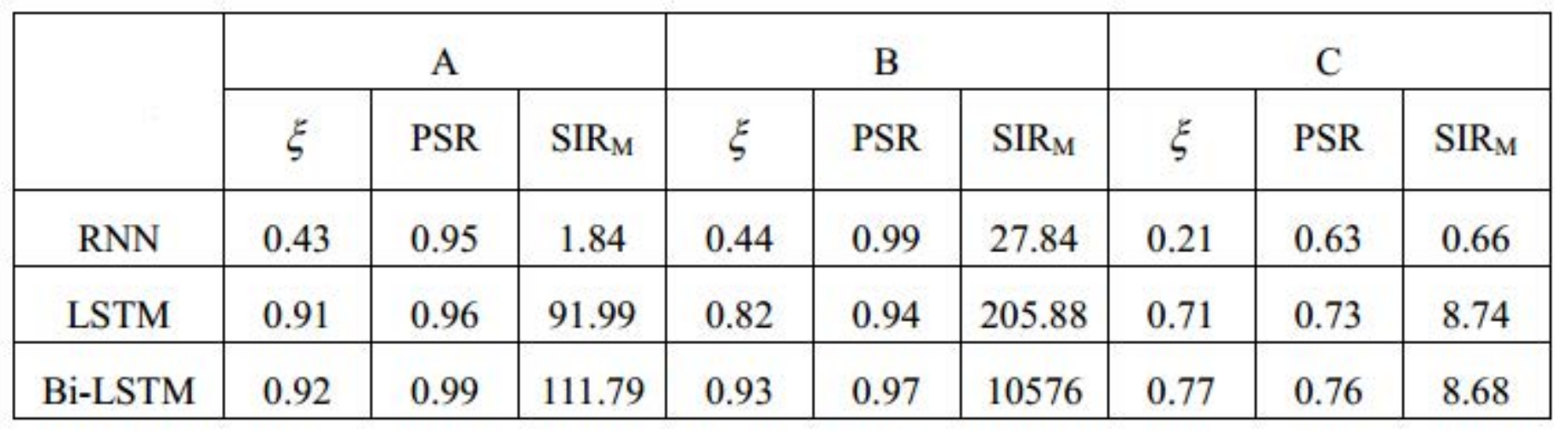}
	\caption{Comparison of RNN, LSTM and Bi-LSTM models}
	\label{three}
\end{figure}

Finally, three models, RNN, LSTM, and Bi-LSTM, were selected for comparison. Each model separates the mixed signals composed of sources A, B and C. The similarity coefficient $\xi$, PSR and $SIR_M$ are selected as comparison indicators, and the comparison results are shown in Fig.\ref{three}.

According to the results of Fig.\ref{three}, the effect of RNN is the worst. Although the PSR of the A and B signals reaches 0.99, their $SIR_M$ is very low. This shows that although the T-F information of the source signal is preserved, most of the T-F points that are not part of the source signal are also classified as source signals. The recovered signal then contains other signal components. In addition to keeping the original information well, LSTM and Bi-LSTM can also implement interference suppression for other signals. Bi-LSTM has a better suppression effect than LSTM. Signal B has the best recovery in the three configurations, especially in Bi-LSTM where the $SIR_M$ even reaches 10576 and the PSR reaches 0.97. Comparing the distribution of the three sources, it can be seen that the signal B and the signals A and C have almost no overlap in the frequency domain, so they are easily distinguished.

\section{Conclusion}
In this paper, a deep learning separation method for underwater acoustic signals based on T-F mask method is proposed. It mainly uses the Bi-LSTM to create the features of Time-Frequency mask for clustering. In this way, every T-F bins is "encoded" directly and partitioned into a reasonable region according to their magnitude. For real-world tasks, it is important for the proposed model to have a good scalability since the number of target sources is not fixed. At the same time, the model should have good generalization ability, so that it can work effectively when separating uncertain underwater acoustic mixing sources in online applications. In order to illustrate the universality and extensibility of the model, we conducted experiments on two unknown mixed sources and three mixed sources respectively, and tested the robustness of the model by adding 0-40dB gaussian noise. Finally, this paper compares and analyzes the performance of RNN, LSTM and Bi-LSTM networks in extracting underwater acoustic signal characteristics. The results show that the proposed method can obtain better performance under the conditions of large mixed signal uncertainty and large Gaussian noise, and it has obvious improvement compared with the traditional T-F mask method. The most important thing is: Compared with mainstream methods, this model not only has better separation performance in binary signal separation, but also can effectively separate aliased signals in the case of multiple signal separation that cannot be well handled by existing methods.

\bibliographystyle{ieeetr}

\setlength{\baselineskip}{10pt}

\bibliography{referencebibtex}

\end{document}